\documentclass[12pt]{article} % For LaTeX2e
\usepackage{authblk}

\usepackage{booktabs}

\usepackage[table,xcdraw]{xcolor}
\usepackage[utf8]{inputenc} % allow utf-8 input
\usepackage[T1]{fontenc}    % use 8-bit T1 fonts
\usepackage{url}            % simple URL typesetting
\usepackage{booktabs}       % professional-quality tables
\usepackage{amsfonts}       % blackboard math symbols
\usepackage{nicefrac}       % compact symbols for 1/2, etc.
\usepackage{microtype}      % microtypography
\usepackage{xcolor}         % colors
\usepackage[colorlinks,citecolor=blue,urlcolor=blue]{hyperref}
\usepackage{wrapfig}

\usepackage{bbm}
\usepackage{booktabs}
\usepackage{amsmath}
\usepackage{amsfonts}
\usepackage{amssymb}
\usepackage{amsthm}
\usepackage{bbm}

\hypersetup{%
  colorlinks = true,
  linkcolor  = blue
}
\usepackage{url}
\usepackage{natbib}

\usepackage{graphicx}
\usepackage{multirow}
\usepackage{amsthm}
\usepackage{algorithm}
\usepackage{algpseudocode}

\theoremstyle{plain}

\theoremstyle{definition}

\title{{\LARGE{\bf Forecasting West Nile virus with deep graph encoders}}}
\author[1]{Ethan Grieffenstein}
\author[2]{Trevor Harris}
\author[3]{Rebecca Smith}
\affil[1]{Department of Statistics, Texas A\&M University}
\affil[2]{Department of Statistics, University of Connecticut}
\affil[3]{Department of Pathobiology, University of Illinois at Urbana-Champaign}

\begin{document}
\maketitle

\begin{abstract}
West Nile virus is a significant, and growing, public health issue in the United States. With no human vaccine, mosquito control programs rely on accurate forecasting to determine when and where WNV will emerge. Recently, spatial Graph neural networks (GNNs) were shown to be a powerful tool for WNV forecasting, significantly improving over traditional methods. Building on this work, we introduce a new GNN variant that linearly connects graph attention layers, allowing us to train much larger models than previously used for WNV forecasting. This architecture specializes general densely connected GNNs so that the model focuses more heavily on local information to prevent over smoothing. To support training large GNNs we compiled a massive new dataset of weather data, land use information, and mosquito trap results across Illinois. Experiments show that our approach significantly outperforms both GNN and classical baselines in both out-of-sample and out-of-graph WNV prediction skill across a variety of scenarios and over all prediction horizons.
\end{abstract}

% \tableofcontents

\section{Introduction}

% \paragraph{What is the problem domain?}
West Nile virus (WNV) is the leading cause of mosquito-borne illness in the United States \citep{thomson2022climate}. Between 1999 and 2019, over 7 million human cases were reported with over 50,000 severe cases of West Nile fever and 2400 fatalities \citep{petersen2013west, ronca2019cumulative, mcdonald2021surveillance, sass2022impact}. WNV was first isolated from a febrile illness case in Uganda \citep{smithburn1940neurotropic}, and has since spread across southern and eastern Europe, west Asia, Australia, and North and South America. The enzootic cycle of WNV is driven by adult mosquito blood feeding, primarily \textit{Culex} mosquitoes in North America, on over 300 susceptible bird species \citep{taylor1956study, thomson2022climate}. However, it can often spill over into human populations and cause severe illness. Over the next century, climate change will continue to expand the range of WNV by reshaping local weather patterns, hydrology, vegetation, and land usage \citep{soverow2009infectious, paz2015climate}. This has already been observed by the emergence of WNV in North America in 1999 \citep{anderson2006importance} and Germany in 2018 \citep{frank2022west}.

% \paragraph{What is the problem specifically?}
% WNV has become a significant and growing issue in the United States and worldwide. 
With no human WNV vaccine, the current best mitigation strategy is to forecast when and where WNV might emerge so that local populations can be forewarned \citep{sampathkumar2003west}. Forecasting models are used to identify and track the spread of WNV, identify high-risk populations, and monitor the effectiveness of control measures. Forecasts also help guide public health policy, such as how to allocate resources for WNV prevention and where to target insecticide programs \citep{bellini2014review}. Forecasting WNV prevalence has, therefore, become a top priority for mosquito control programs \citep{benelli2015research}. 
However, directly modeling WNV cases is often prohibitive due to data scarcity and privacy concerns using health record data. Instead, forecasters model the local mosquito infection rates (IR) during mosquito season (May to October) as a proxy for WNV threat. This approach is commonly used in developing early warning systems \citep{defelice2017ensemble} due to the abundance of mosquito infection data collected via extensive mosquito surveillance programs \citep{engler2013european}.

% then analyzing the product of these two quantities, the vector index. The vector index neatly summarizes the overall risk posed by WNV in a given region and is an established mediator of WNV case rates.
% We will focus on modeling both aspects of the vector index, namely abundance and IR, as a surrogate for human case modeling.

There are a wide variety of environmental factors that effect WNV infection rates in mosquitoes, including local weather conditions, hydrological factors, vegetation levels, and land usage patterns \citep{winters2008combining, soverow2009infectious, rochlin2019west, karki2020drivers}. Historically, temperature and precipitation were thought to be the greatest drivers of WNV \citep{paz2013environmental}. Higher temperatures lead to faster mosquito lifecycles, more frequent blood feeding, and increased virus amplification \citep{hartley2012effects, reisen2013ecology}. 
% However, too high (above $30^\circ$C) temperatures can lead higher larval and pupa mortality rates, yet brief extreme weather, such as heat waves, are associated with emergence and reemergence of WNV (citation).
Precipitation is necessary for creating stagnant water sources for egg laying, primarily in sewers and tires \citep{landesman2007inter}, although, conversely, extreme precipitation can dramatically increase egg, larval, and pupa mortality by flushing them out of the local watershed.
Weather effects can be further complicated by local geographic and environmental factors \citep{bowden2011regional}. The structure of the local sewershed can modulate the effects of precipitation by preventing or amplifying flushing effects \citep{koenraadt2008flushing}. Land use patterns, such as increases in irrigated cropland, fragmented forests, and urbanization have also lead to increases in WNV cases \citep{marcantonio2015identifying}.

% \paragraph{Why are existing methods no good?}
Although WNV is a complex, interacting, systems of factors, many forecasting approaches still assume simple linear relationships and autoregressive models on highly aggregated features \citep{barker2019models, poh2019influence, keyel2021proposed, holcomb2023evaluation}. These models also, typically, impose strong parametric assumptions on the outcome variable, such as a negative-binomial or gamma distribution.
% statistical modeling approaches revolve around simple linear models on high aggregated features. Models such as (citation) assume a linear form based on aggregate features, such as weekly temperature and precipitation means. (another example here).
More flexible approaches improve on these models by accounting for spatial factors and spatial correlation \citep{uelmen2020effects, uelmen2021dynamics} or by utilizing more complex feature representations based on functional data \citep{sass2022impact}. Recently, complex machine learning forecasting systems \citep{farooq2022artificial, tonks2024forecasting} have emerged as a powerful alternative to classical approaches.

Despite recent advances in forecasting skill, these models are limited in their ability to process high resolution spatio-temporal data, which has been instrumental in scaling prediction systems in other domains \citep{sun2019deep, greenspan2009super}. The graph neural network (GNN) \citep{wu2020comprehensive, zhou2020graph} model introduced in \cite{tonks2024forecasting} is potentially scalable due to its inductive learning framework: GraphSAGE \citep{hamilton2017inductive}. GraphSAGE is computationally efficient, but, unfortunately, is also provably degenerate with increasing depth \citep{li2018deeper}. This limits its ability to process large-scale spatiotemporal data, where deep models are dominant \citep{wu2020comprehensive}, and lead to \cite{tonks2024forecasting} using a small, shallow GNN model on aggregate features. However, even with these limitations, their model still significantly out-performed simple linear and spatial models on WNV forecasting in Illinois across multiple time horizons.
% \cite{tonks2024forecasting} introduced a graph neural network \citep{wu2020comprehensive, zhou2020graph} model to handle complex non-linear relationships while accounting for spatial dependence. This model used the popular GraphSAGE framework for inductive graph learning \citep{hamilton2017inductive}. It is computationally efficient, but has limited scalability due to degeneracy issues with increasing depth \citep{li2018deeper}. This limits its ability to process high resolution spatiotemporal information with very deep models, so, instead \cite{tonks2024forecasting} used simple aggregate features, such as a single temperature value for all nodes, and a small, shallow GNN.

% \paragraph{What are we proposing?}

To train much larger GNN models for WNV forecasting, we propose a modified GNN architecture based on densely connected graph layers \cite{guo2019densely}. Densely connected GNNs introduce skip connections \citep{he2016deep} between graph layers, as in densely connected convolutional neural networks \citep{huang2017densely}, which allows information to flow through the GNN without necessarily being processed by each layer. This was shown to significantly improve the performance of GNNs by stabilizing the training dynamics of deep GNNs and preventing over-smoothing due to repeated graph convolution. In our proposed architecture, instead of connecting all layers to the output, we connect the input to all intermediate layers, which allows the self-information at each node to be refreshed after each message-passing operation. 
% Furthermore, we introduce a new convolutional attention pooling mechanism that efficiently fuses multi-modal information across data modes and propagates the dominant signals to neighboring nodes.
We show that this simple architectural change allows us train massive GNNs on much larger datasets with far more complex covariate information than pervious works \citep{tonks2024forecasting}. In accordance with widely observed empirical scaling laws for deep neural networks \citep{hestness2017deep,kaplan2020scaling,sorscher2022beyond}, we show that this leads to significantly more accurate and robust predictions, and thus greatly improved early warning skill \citep{holcomb2023evaluation}. 

% Contributions: New dataset, new GNN architecture, vaslty improved results on important problem

% This simple architectural change allows us train significantly larger GNNs on much larger and more complex datasets. 

% We propose a new GNN model, based on densely connected graph layers (citation), that allows us to massively scale GNNs for spatiotemporal modeling problems, such as WNV forecasting. Instead of  

% Following the work of (citation), we propose a massively scalable graph neural network model based on convolutional attention (CAT) that allows for efficient processing of node-wise spatiotemporal information. In our proposed model, multi-modal spatiotemporal covariates will be associated with each node (trap site) in the spatial graph and a new convolutional attention pooling aggregator will be used to fuse information across data modes and propagate the dominant signals to neighboring nodes. Thus our model will be able to handle far more complex covariate information than (citation).

% \paragraph{How is this novel?}
% We use a deep learning graph neural network architecture to handle large and complex weather and trap data. Semi-supervised learning is leveraged to share maximal information between nearby nodes. Our method out-performs other approaches applied in the West Nile virus forecasting space. 

\section{Dataset} \label{sec:Dataset}

\paragraph{Mosquito Traps} Target data for modeling mosquito infection rates comes from surveillance programs, such as the North Shore Mosquito Abatement District in Chicago, IL. Surveillance programs place and monitor mosquito traps \citep{nsmad} around the county for the presence of WNV.
Mosquito trap data was provided by the Illinois Department of Public Health (IDPH) and includes 133,867 mosquito trap results, across 333 unique traps, from 2008 through 2021. Each observation records whether the mosquitoes in that trap on that day tested positive for WNV. This dataset also includes trap specific information such as latitude, longitude, sample collection date, and the number of mosquitoes tested (pool size). 
Most trap locations are concentrated around Chicago and other major population centers (Champaign, Peoria, and Carbondale), however, the dataset also includes observations from 94 of Illinois' 102 counties. Thus, our data represents a wide range of urban and rural environments.

To account for spatial correlation, and meaningfully apply a graph neural network (GNN) model, we connected each trap with its $k$-nearest neighbors based on geographic proximity. Spatially linking traps allows the GNN to share covariate information across trap sites, which has been shown to improve predictive skill \citep{uelmen2020effects, tonks2022forecasting}. In addition to the provided covariates, we augmented each node (trap site) with additional covariates based on nearby local historical meteorological, environmental, and hydrological characteristics.
% To account for spatial variations in environmental and mosquito activity patterns, we further linked each trap with its k-nearest neighbors based on geographic proximity. This spatial linking is important because it captures the influence of nearby environmental conditions and mosquito populations, (which are correlated across space maybe??). 
% A long line of work has established a strong connection between local environmental factors and the emerging presence of WNV in a region.
% We will augment our mosquito trap dataset by attaching local historical meteorological, environmental, and hydrological characteristics to each observation to improve our model's ability to forecast near-future trap positivity. 
This data was sourced from high resolution weather \citep{di2008constructing} and land use maps \citep{homer2012national}.

% Our dataset is a compilation of historical climate data and environmental characteristics surrounding mosquito trap locations, primarily focused in Chicago, Illinois, with additional traps positioned throughout the state. This collection includes 333 unique traps, representing a wide range of urban and rural environments, this allows for the understanding of factors influencing mosquito populations and the rate of West Nile virus in the area. Dataset covers 13 years of data, from 2008 to 2021, giving us ample historical context for examining how fluctuations in climate variables such as temperature and precipitation and correlate with mosquito activity. 

% \subsection{Data Sources}

% The dataset is constructed from two primary sources:

\paragraph{PRISM Climate Data} 

\begin{figure}[htbp]
\centering
\includegraphics[trim={5cm 0cm 5.95cm 0cm}, clip, width=0.235\textwidth]{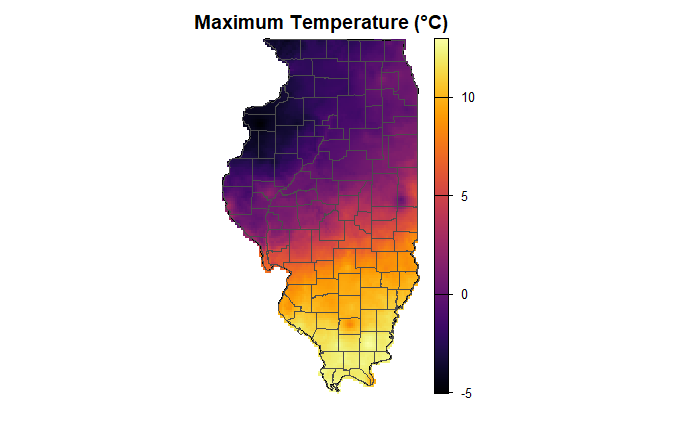}
\includegraphics[trim={5cm 0cm 6cm 0cm}, clip, width=0.235\textwidth]{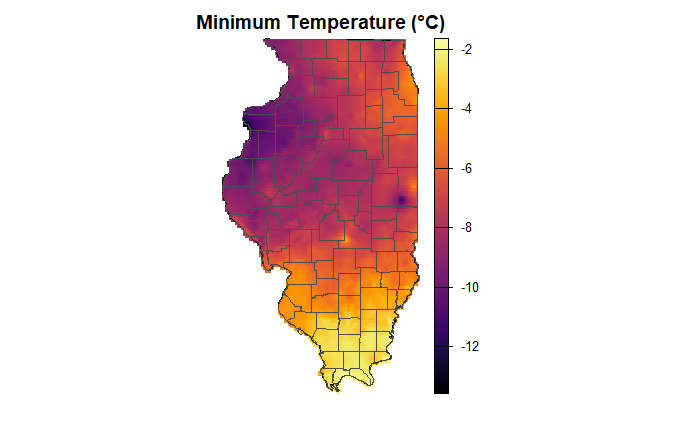}
\includegraphics[trim={5cm 0cm 5.95cm 0cm}, clip, width=0.235\textwidth]{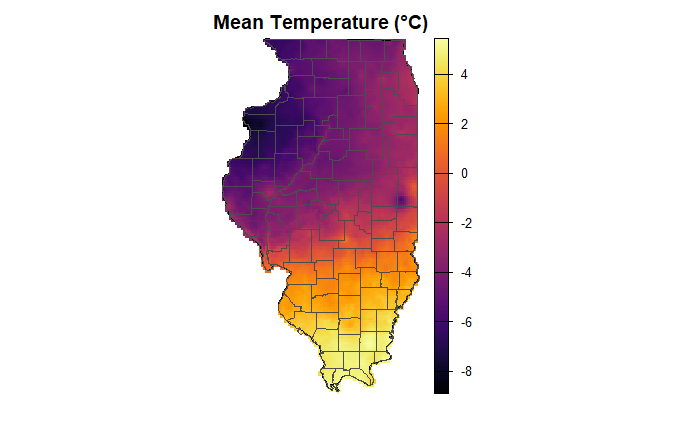}
\includegraphics[trim={5cm 0cm 5.95cm 0cm}, clip, width=0.235\textwidth]{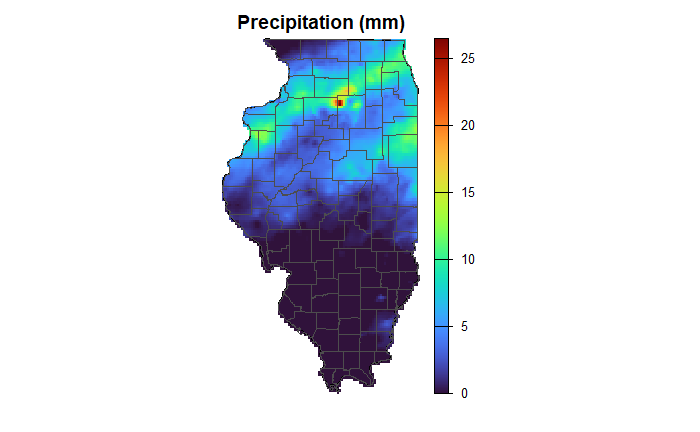}

\caption{Four remotely sensed Prism quantities at 4km grid resolution. Observations are present in 48 out of 102 counties, with a high concentration in Cook County.}
\label{fig:NLCD_Data}
\end{figure}

The PRISM climate data is a high resolution (4km grid scale) synthesis of climate observations from a wide range of monitoring networks \citep{Daly2008}. PRISM provides quality controlled daily measurements of key meteorological features such as daily minimum, mean, and max temperature, and daily total precipitation.
% covers the continental U.S. at a spatial resolution of 4 km, providing daily records of key climate variables from 
% 2008 to 2021
% These variables include:
% \begin{itemize}
%     \item Daily Mean Temperature
%     \item Daily Maximum Temperature
%     \item Daily Minimum Temperature
%     \item Daily Precipitation
% \end{itemize}
PRISM records offer a high degree of temporal granularity (daily), enables the analysis of weather events such as significant shifts in historical temperature and precipitation. 
From temperature and precipitation measurements, we can derive the commonly used heating and cooling degree days \citep{reisen2006effects, shocket2020transmission}.
% In addition to temperature and precipitation, heating and cooling degree days have been derived to capture historical cooling and heating.
A heating degree day is defined as any day where the mean temperature exceeds 65°F (18.3°C), while a cooling degree day occurs when the mean temperature falls below this threshold. 
% Prior studies have show degree days to be important in the forecasting of WNV.
Prior studies have shown that degree days are significantly more predictive than raw climatic variation \citep{reisen2006effects}.
% , a finding we also observed.

\paragraph{National Land Cover Database (NLCD)}

\begin{figure}[htbp]
\centering
\includegraphics[trim={5.5cm 0cm 6cm 0cm}, clip, width=0.235\textwidth]{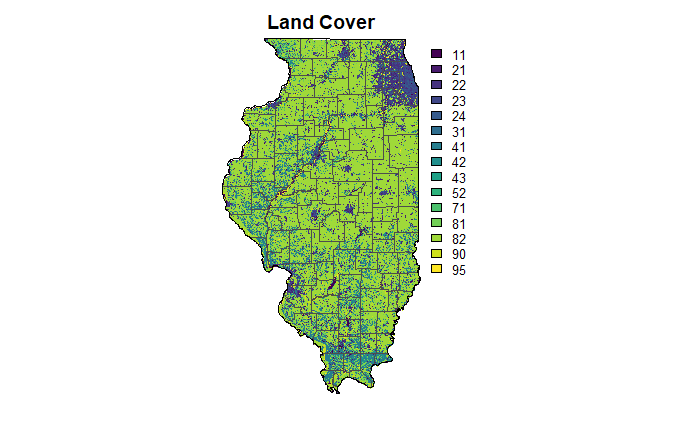}
\includegraphics[trim={5.5cm 0cm 5.95cm 0cm}, clip, width=0.235\textwidth]{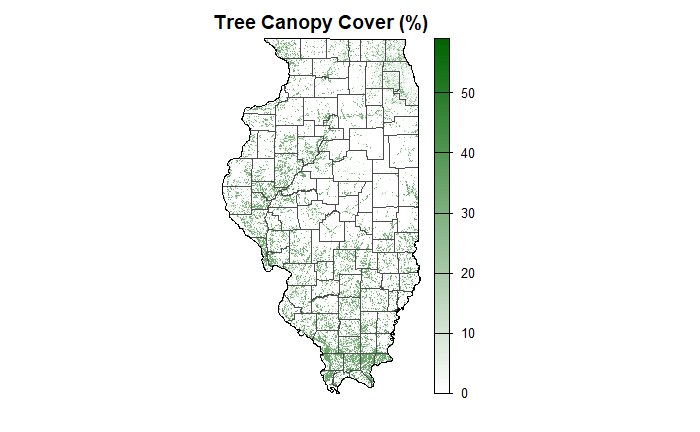}
\includegraphics[trim={5.5cm 0cm 6cm 0cm}, clip, width=0.235\textwidth]{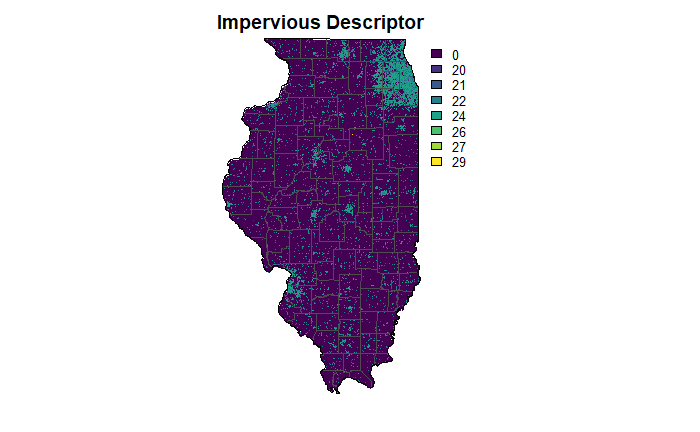}
\includegraphics[trim={5.5cm 0cm 5.95cm 0cm}, clip, width=0.235\textwidth]{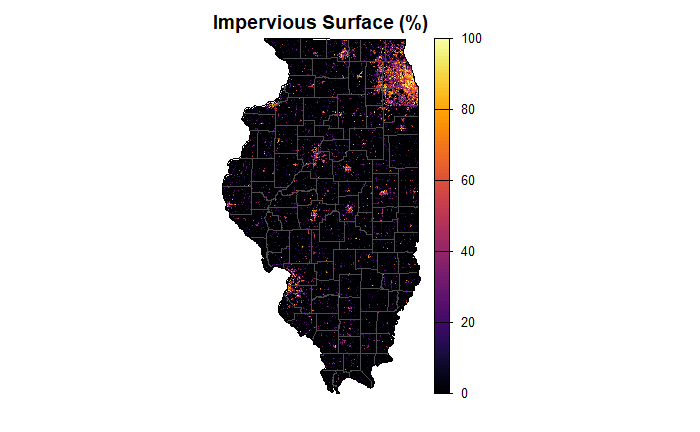}

\caption{Four remotely sensed NLCD quantities at 30m resolution. Observations are present in 48 out of 102 counties, with a high concentration in Cook County.}
\label{fig:NLCD_Data}
\end{figure}

The National Land Cover Database (NLCD), created by the U.S. Geological Survey (USGS) in collaboration with the Multi-Resolution Land Characteristics (MRLC) Consortium, integrates a wide variety of remotely sensed data products and is one of the most widely used datasets for geospatial analysis in the United States \citep{homer2012national}. For this study, we utilize NLCD variables including urban imperviousness, urban impervious descriptors, land cover classifications, and tree canopy data, all at a 30 m pixel resolution. Urban imperviousness quantifies the percentage of impervious surfaces (e.g., roads, buildings) within each developed pixel, while urban impervious descriptors provide detailed classifications of these surfaces, such as road types and urban area categories. Land cover data classifies each pixel into one of sixteen possible land types, such as forests, wetlands, or grasslands, while tree canopy data represents the percentage of tree cover at each location and is available for the years 2011 to 2021. These variables are closely related to environmental factors that can predict West Nile Virus (WNV) activity \citep{karki2020drivers}.

\paragraph{Data Processing} After extraction, we transformed the data from continuous summary statistics into interpretable categorical and binary indicators for modeling purposes.
For tree canopy, we categorized coverage values into three quantiles: areas with <20 percent coverage were labeled as ``low'', those between 20 percent and 50 percent as ``medium'' and those above 50 percent as ``high''. Similarly, for impervious surfaces, areas with less than 33 percent coverage were categorized as ``low,'', those between 33 percent and 67 percent as ``medium'', and those above 67 percent as ``high''.
For land cover, we assigned a binary value of 1 to indicate the presence of a land cover type if it accounted for more than 15 percent of the area around a trap. This approach allowed for the identification of multiple land cover types when more than one exceeded the 15 percent threshold.
The impervious surface descriptors were handled in a similar manner: binary values were assigned to indicate the presence of specific road features (primary road, secondary road, tertiary road, non-road impervious surfaces) if the corresponding feature covered more than 10 percent of the area. 
Daily temperature data collected from PRSIM was summarized using the 10th, 25th, 50th, 75th, and 90th percent quantiles to capture the range the climate conditions. 
While this pre-processing reduces the dimensionality by collapsing the data into compact forms, the resulting dimension is still high dimensional and more complex than what classical methods can typically handle.. 
% \textbf{(Some how flow into this better as I thinks its important to bring up that poolsize was used)} 

% We will merge the PRISM and NLCD datasets with the primary mosquito trap dataset by spatially linking the environmental data to the specific locations of each trap. For each trap, PRISM and NLCD data were collected at the corresponding geographic coordinates to capture localized climate and land characteristics. 

\section{Method} \label{sec:method}

Let $Y(s, t) \in \mathcal{Y} = \{0, 1\}$ denote the test result measured at spatial location $s \in D \subset \mathbb{R}^2$ and time point $t \in \mathcal{T} \subset \mathbb{R}$. Without loss of generality we assume $D = [0, 1]^2$ and $\mathcal{T} = [0, 1]$. Let $X(s, t) \in \mathcal{X}$ denote the covariate information recorded at, or around, spatial location $s$ and time point $t$. For each time $t \in \mathcal{T}$, we define a spatial graph $G_t = (V_t, E_t)$, where $V_t$ denotes the vertex, or node, set and $E_t$ denotes the edge set at time $t$. 
Mosquito traps become nodes and we link them with (truncated) kNN to form spatial graphs.
For a spatial graph, the node set $V_t$ is the set of all locations $s \in D$, where the pair $(X(s, t), Y(s, t))$ is observed at time $t$. The edge set $E_t$ can be defined in many different ways, however, we will follow \cite{tonks2024forecasting} and define $E_t$ by the $k$-nearest neighbor distances. That is, $v_1 \in V_t$ is connected with $v_2 \in V_t$ if the distance between their respective spatial locations, $d(s_1,  s_2)$, is among the $k$ smallest distances $d(s_j, s_2)$ for all nodes $v_j \in V_t$. Here $d(\cdot, \cdot)$ represents an arbitrary distance notion between two points, such as Euclidean distance. We denote the space of all possible graphs $G_t$ with up to $n$ nodes as $\mathcal{G}_n$.

% \paragraph{Graph neural networks}
A graph neural network is a function $f_\theta : \mathcal{G} \times \mathcal{X} \mapsto \mathcal{Y}$ that maps a graph $G \in \mathcal{G}_n$ and the node covariates $X(s, t)$ to the node response $Y(s, t)$ for all nodes in $G$. In our application, given a trap location $(s_0, t_0)$, a graph neural network would map the covariates $X(s_0, t_0)$, and the covariate information of neighboring traps $X(s_1, t_1),..., X(s_m, t_m)$ for $m \leq n$, to the result $Y(s_0, t_0)$. We use $\theta$ to represent all parameters, i.e. weights and biases, of the GNN $f_\theta$. 
There are a wide variety of GNN architectures for constructing the function $f_\theta$, including graph convolutional networks (GCN), graph attention networks (GAT), GraphSAGE, and others \citep{guo2019densely, velivckovic2018graph, hamilton2017inductive}. Because we observe a sequence of graphs $G_t$ for $t \in \mathcal{T}$ that evolve over time, we base our approach on an \textit{inductive} architecture called GraphSAGE, that learns a function for embedding nodes based on node features and the local graph topology. Inductive approaches naturally generalize to new nodes and new graphs because they are not reliant on any single graph. 

% We denote our approach GraphDense.

\subsection{Model Architecture} \label{sec:model_architecture}

Graph neural networks have long been noted to scale poorly with depth \citep{li2019deepgcns, li2018deeper, zhou2020graph}, which historically limited GNNs to relatively shallow architectures as in \cite{tonks2024forecasting}. Not only does gradient based training become increasingly unstable with increasing depth \citep{li2019deepgcns}, as in ordinary feed forward neural networks \citep{veit2016residual, balduzzi2017shattered}, deep graph convolutional network (GCNs) provably collapse all nodes to the same value with increasing depth \citep{li2018deeper}. This latter fact is unique to graph convolution, a type of iterated graph Laplacian smoothing, and is often observed in practice even after only a few layers \citep{li2019deepgcns, guo2019densely}.

Recent works have tried to circumvent gradient and depth based degeneracy issues through architectural modifications, such as Column Network (CLN) \citep{pham2017column}, Highway GCNs \citep{rahimi2018semi}, and skip connections \citep{li2019deepgcns, guo2019densely}. The ResGCN architecture \citep{guo2019densely} is a particularly successful and computational efficient scheme for enabling deep GNN training. ResGCN, and the more general DenseGCN, augment a standard deep GCN with dense connections between graphs layers, and dialated convolutions, to improve the flow of information through the network and efficiently reuse learned feature maps. This idea stems from earlier work on densely connected Convolutional Neural Networks (CNN) \citep{huang2017densely}, which were shown to improve performance in image classification tasks.

% \begin{figure}[!ht]
%     \centering
%     \includegraphics[width=0.5\textwidth]{figs/ModelFigs/GraphMAGEDiagram.png} 
%     \caption{Overview of the Mage Dense Architecture from three perspectives.}
%     \label{fig:GraphMAGEThreePanel}
% \end{figure}

\begin{wrapfigure}{R}{0.5\textwidth}
    \centering
    \includegraphics[width=0.99\linewidth]{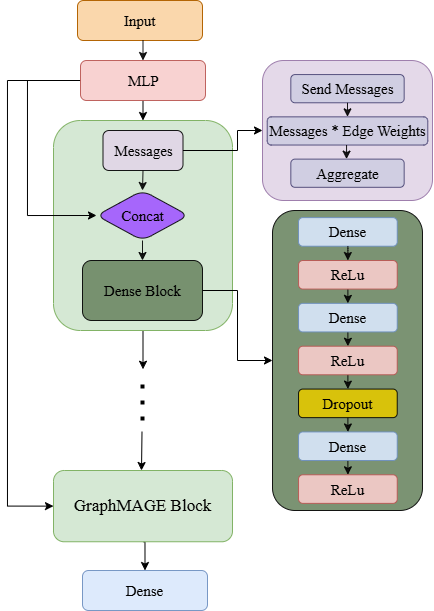} 
    \vspace{-1em}
    \caption{Overview of the GraphMAGE architecture, highlighting the flow of MLP output into subsequent layers.}
    \label{fig:GraphMAGEThreePanel}
    \vspace{-1em}
\end{wrapfigure}

Although ResGCN improves the stability and generalization skill of deep GCN models, it still may suffer from depth saturation due to the exponentially large number of sub-models with a high depth implied by the residual connections \citep{veit2016residual}. Furthermore, it is also reliant on dilated convolutions, which introduces further hyperparameters. Therefore, we introduce a restricted form of ResGCN called GraphMAGE (Figure \ref{fig:GraphMAGEThreePanel}) that removes the need for dilated convolutions and allows for more efficient deep scaling. GraphMAGE only includes skip connections from the embedding layer to the intermediate layers, without any residual connections in between. Intuitively, we are ``refreshing'' the self-information at each node after each message passing operation to prevent over-smoothing from neighboring the nodes. 

We define an $L$-layer GraphMAGE model (Figure \ref{fig:GraphMAGEThreePanel}) as an inductive embedding algorithm (Algorithm \ref{alg:GraphMAGE}). GraphMAGE embeds a node $v \in V$ at the $l$-th layer as
\begin{equation}
    \begin{aligned}
        h^l_{N(v)} &\leftarrow \text{AGG}(h^{l-1}_u, \space \forall u \in N(v)) \\
        h^l_v &\leftarrow  \text{MLP}(\text{Concat}(h^0_v, h^l_{N(v)}))
    \end{aligned}
\end{equation}
where $h^0_v$ is the initial embedding of node covariates $X(s_v, t)$, $h^{l-1}_v$  is the previous layers embedding, MLP is a multi-layer perceptron model, $N(v) \subset V$ are the neighbors of $v \in V$, and AGG is a user-defined aggregation function for combining neighbor representations $h^{l-1}_u$ $\forall u \in N(v)$. Common choices for AGG include means, pooling, and LSTMs \citep{hamilton2017inductive}. Essentially, at each layer, GraphMAGE aggregates the information surrounding node $v \in V$ into a latent representation $h^l_{N(v)}$. This is then combined with the initial latent representation $h^0_v$ via an MLP, to get the updated state $h^l_v$ for node $v \in V$.
The full embedding algorithm is defined in Algorithm \ref{alg:GraphMAGE}.
\begin{algorithm}
	\caption{GraphMAGE} \label{alg:GraphMAGE}
	\begin{algorithmic}
        \Require $G(V, E)$, input features $\{x_v : v \in V \}$, Depth $L$ and multi-layer perceptrons MLP$_1$,..,MLP$_L$, Differentiable AGG function
		\For {$l=1,\ldots,L$}
			\For {$v \in V$}
				\State $h^l_{N(v)} \leftarrow \text{AGG}(h^{l-1}_u, \space \forall u \in N(v))$
				\State $h^l_v \leftarrow  \text{MLP}_l(\text{Concat}(h^0_v, h^l_{N(v)}))$
			\EndFor
			% \State Normalize?
			% \State Anything else?
		\EndFor
	\end{algorithmic} 
\end{algorithm}

Algorthm \ref{alg:GraphMAGE} is a modification of the original GraphSAGE algorithm, where, instead of concatenating (Concat$(\cdot)$) the processed self-information $h_v^{l-1}$ with the neighbor representation $h^l_{N(v)}$, we concatenate the original, embedded, self-information $h_v^0$. By continually re-adding the (nearly) unprocessed covariates, $h_v^0$, we prevent the model from degenerating with increasing depth. This is because $h_v^0$ is never ``contaminated'' via message passing, and thus can not be diluted.

\subsection{Semi-supervised and fully supervised training} \label{sec:training}

% \begin{figure}[!ht]
%     \centering
%     \includegraphics[height = 3cm,width=0.4\linewidth]{figs/ToyExample.png}  % Reduced width
%     \caption{Example of graph connections}
%     \label{fig:Semi-Supervised example}
% \end{figure}

As with other inductive graph neural network frameworks \citep{hamilton2017inductive, velivckovic2017graph, guo2019densely}, we can train proposed model using either a fully supervised or a semi-supervised learning framework. We detail both frameworks below and how we construct the dynamic graphs and loss function for each.

\paragraph{Supervised training}
A supervised graph $G_s = (V_s, E_s)$ consists of only the nodes that were checked during a given time period \textit{t}. For each node $v_i \in V_s$, there exists an observed outcome $Y(s_i, t_i)$ and associated covariates $X(s_i, t_i)$. Edges between nodes $v_i$ and $v_j$ are only formed if both nodes were checked, i.e., if there exists recorded test results for both $(s_i, t_i)$ and $(s_j, t_j)$ and if the geographic distance between their respective locations falls within the $k$-smallest distances. Each $e_{ij} \in E_s$ is associated with a weight or attribute denoting the geographic distance between the connected nodes $v_i$ and $v_j$, allowing our model to account for spatial proximity. The supervised model optimizes a weighted binary cross-entropy loss, where each class is inversely weighted by its proportion in the training data to address the class imbalance.

% \paragraph{Semi-Supervised Graph Learning ($G_{us}$)}
\paragraph{Semi-supervised training}
A semi-supervised graph $G_{us} = (V_{us}, E_{us})$ contains both checked and unchecked nodes. Unchecked nodes represent known testing locations where mosquito abundance or infection data was not recorded during a given time period $t$, but covariate information $X(s,t)$, such as weather or environmental characteristics, is still available. The node set therefore includes all locations, regardless of whether the outcome $Y(s,t)$ is observed. In this case, edges are formed between both checked and unchecked nodes, allowing information to propagate from both checked and unchecked nodes. The edge connections in the semi-supervised graph are constructed similarly to those in the supervised graph, with edges formed based on the geographic proximity of the $k$-nearest neighbors and each edge is weighted with a weight attribute that which allows the spatial proximity of nodes be to accounted for both checked and unchecked nodes. Similarly to supervised training in semi-supervised training we optimize on a weighted binary cross entropy loss, where each class is inversely weighted by its proportion in the training data. 

The semi-supervised graph enables information sharing even in scenarios where a node has very few checked neighbors. By incorporating unchecked nodes with covariate information related to the surrounding climate, the model can propagate knowledge from distant checked nodes and leverage nearby unchecked nodes to help classify locations where close observations are sparse. This is particularly useful for predicting outcomes in regions with limited observations where covariate data can still provide valuable context for forecasting. When the covariates exhibit structure,  i.e. reside on a low dimensional manifold, then semi-supervised training can significantly improve predictive accuracy \citep{wasserman2007statistical}.

% \paragraph{Big models}
% We will present a few variations of our model to show case its strengths. These variations are focused on how we handle the trap specific data versus weather and trap surroundings characteristic data.

% - here talk about some specific implementations: splitting the node specific data or not, supervised vs unsupervised pro and cons
% - talk about poolsize here maybe: mention its pros and cons and show how the model still has good performance on semi supervised graphs without relying heavily on poolsize thus maybe this leads to the application of prediction across a large area given the long  term historical weather data and area characteristics 
% - Talk about training regime tricks that we used to overcome a sticky issue encountered early on 
% - Include images

% \begin{itemize}
%     \item Baseline GraphSage Model
%     \item Graph Dense with Semi-Supervised Graphs 
%     \item Graph Dense with Supervised Graphs
%     \item Random Forest
% \end{itemize}

\subsection{GraphMAGE vs ResGCN}

The GraphMAGE algorithm is special case of ResGCN and, therefore, inherits many of the benefits of residuals connections such as a smoothed loss landscape and non-shattered gradients \citep{li2018visualizing, balduzzi2017shattered}, which enable smoother training of deep models. 
However, our connection strategy dramatically changes the ensemble implied by the residual paths. 

\cite{veit2016residual} showed that residual networks can be interpreted as an exponentially large ensemble of sub-networks of varying path lengths. Furthermore, they showed that the shortest paths played the largest role in forming predictions.
By only connecting the initial embedding to each subsequent layer, we reduce the ensemble size from exponential $O(2^n)$ to linear $O(n)$ in the number of GNN layers. Moreover, this has the effect of flattening the path length distribution from approximately normal, and tightly concentrated around the mean, to exactly uniform. That is we have exactly one sub-network per path length meaning GraphMAGE behaves like an ensemble of distinct $1,2,3,...,L$-layer GraphSAGE models.

Flattening the path length distribution is important for graph models, which provably collapse with increasing depth \citep{li2018deeper}, by ensuring there are an equal number of short path (shallow) models to counterbalance the long path (deep) models. Using the full ResGCN would result in an ensemble of GNNs with depths mostly concentrated around $L/2$. If $L$ is large, then these models may be severely degenerate, yet still out-weigh the shallow models due to sheer volume. GraphMAGE, on the other hand, always has a uniform path length distribution, meaning that it still considers very long range connections, but, by construction, always weights nearby neighbors highly.

\section{Application to WNV forecasting} \label{sec:Application}

We evaluate our proposed GraphMAGE models on a large-scale WNV forecasting problem in Illinois. We will use our new dataset (Section \ref{sec:Dataset}) to train fully supervised (SL) and semi-supervised (SSL) deep GraphMAGE models to forecast the presence of WNV at all trap locations in Illinois. We focus on the Chicago area where seasonal climate variability, diverse land-use characteristics, and relatively large sample sizes allow us to evaluate performance across a wide range of settings. This includes urban-rural divides where the graph topology, missingness, control strategies, and overall WNV risk can vary significantly.

We construct our model (Figure \ref{fig:GraphMAGEThreePanel}) by stacking four GraphMAGE layers (Width = 128), as described in Sections \ref{sec:method}, using relu activations and Dropout (p = 0.2). We will consider both a supervised (GraphMAGE-SL) and an semi-supervised (GraphMAGE-SSL) variant of our approach (Section \ref{sec:training}). Unless stated otherwise, we train on all traps in the first 8 years of data (29838 obs) and evaluate on all traps in the last 3 years of data (10752 obs). Traps are connected to their (at most) 10 nearest neighbors within a radius of 50 km. We optimize our models using the Adam optimizer with weight decay \citep{kingma2014adam} with adaptive gradient clipping \citep{brock2021high} and cosine decay learning rate scheduling (Section \ref{adx:sec:implementation}). All computation is carried out on a single NVIDIA 2060 GPU.

\subsection{Baselines and metrics}
Baseline approaches for near term WNV forecasting include the GraphSAGE model \cite{tonks2024forecasting}, logistic regression (LR) \citep{uelmen2021dynamics}, and random forest (RF) \citep{breiman2001random}. We recreate the GraphSAGE model in \cite{tonks2024forecasting} using four layers, with width eight, and relu activations, exactly as in the original paper which we will refer to as MageNet Baseline. We train their model with weighted Adam optimizer using the same setting, data and spatial graph as our proposed approach. We construct the RF with 1000 trees, with balanced class weights, and LR with weighted Adam optimizer. The RF and LR are given the same covariate information as the graph models but treat each trap as independent (no information sharing). We additionally include comparisons with ResGCN, using our model and GraphSAGE as the base, to isolate improvements due to our proposed connection strategy.

We consider three primary metrics to evaluate each approach: F1 Score, AUC, and Sensitivity \citep{powers2020evaluation, holcomb2023evaluation}. Additional metrics (Brier score, Accuracy, and Specificity) are included in the appendix (Table \ref{tab:metrics_trainedOnAll_overall_appendix}). The F1 score measures how well each classifier separates positive and negative classes, AUC is a threshold independent measure of false-positive and false-negative tradeoff, and Sensitivity measures the ability to correctly detect positive classes. The F1 score and sensitivity are defined as
% \begin{equation}
% \begin{minipage}{0.45\linewidth}
% \[
% \text{F1 Score} = \frac{\text{TP}}{\text{TP} + \frac{1}{2}(\text{FP} + \text{FN})}
% \]
% \end{minipage}\hfill
% \begin{minipage}{0.45\linewidth}
% \[
% \text{Sensitivity} = \frac{\text{TP}}{\text{TP} + \text{FN}}
% \]
% \end{minipage}
% \end{equation}
\begin{equation}
    \text{F1 Score} = \frac{\text{TP}}{\text{TP} + \frac{1}{2}(\text{FP} + \text{FN})} \quad\quad\quad \text{Sensitivity} = \frac{\text{TP}}{\text{TP} + \text{FN}},
\end{equation}
where TP denotes the true positive rate, FP the false positive rate, and FN the false negative rate. The AUC is defined as the area under the true positive vs false positive curve for all possible classification thresholds \citep{holcomb2023evaluation}.

In all cases, we want all metrics to be higher (closer to 1). Sensitivity is particularly important due to the high cost of missing a positive trap and letting WNV go undetected.
We evaluate each method on its ability to predict trap positivity from one to eight weeks into the future. Because mosquitoes have a natural life cycle of about four weeks, high forecasting skill beyond four weeks is challenging \citep{tonks2024forecasting}.

\subsection{Forecasting skill} \label{sec:forecasting_skill}
We first train all models on all data between 2008 and 2015 and evaluate them on all data between 2016 and 2018. Due to randomness in the initialization of each model, we train each model 20 times using different initializations. The results in Table \ref{tab:metrics_trainedOnAll_overall} show the forecast metrics averaged over these 20 runs.

The F1 scores, Sensitivities, and AUC scores in Table \ref{tab:metrics_trainedOnAll_overall} show that both variants of our proposed model (GraphMAGE-SL and GraphMAGE-SSL) significantly improve over existing approaches across all prediction horizons (1-8 weeks). Both variants have higher F1 scores, Sensitivities, and AUC scores than all other baselines. In particular, our model shows a dramatic improvement over the GraphSAGE based model (around 20\% higher Sensitivity and 40\% higher F1 score), indicating that performance improvements are not merely stemming from using a nonlinear model or spatial connectivity. The ResGCN connection performs well, however, our model achieves a sizable improvement across all metrics except sensitivity. Although ResGCN shows higher sensitivity, this advantage is counteracted by its lower specificity, as shown in the appendix (Table \ref{tab:metrics_trainedOnAll_overall_appendix}). There is little to no difference between GraphMAGE-SL and GraphMAGE-SSL, however, indicating that purely supervised training on evolving graphs may already be performing a type of semi-supervised learning.

% - These results are combined of 20 random seeds

% - still need to get the new random forest numbers

\begin{table}[h]
\centering
\resizebox{\columnwidth}{!}{
\begin{tabular}{l|cccccccc}
\toprule
\multicolumn{9}{c}{\textbf{AUC Score ($\uparrow$)}} \\
\midrule
Model & Week 0 & Week 1 & Week 2 & Week 3 & Week 4 & Week 5 & Week 6 & Week 7 \\
\midrule
GraphMAGE-SSL & \textbf{0.8516} & \textbf{0.8177} & \textbf{0.8089} & \textbf{0.8047} & \textbf{0.8026} & \textbf{0.813} & \textbf{0.8234} & \textbf{0.845} \\
GraphMAGE-SL & 0.8508 & 0.8164 & \textbf{0.8089} & 0.8018 & 0.7963 & 0.8056 & 0.8156 & 0.8443 \\
MageNet Baseline & 0.8059 & 0.7812 & 0.781 & 0.7747 & 0.7782 & 0.7828 & 0.7954 & 0.8147 \\
MageNet Baseline ResGCN & 0.7322 & 0.7074 & 0.6977 & 0.6664 & 0.6950 & 0.6693 & 0.7180 & 0.7379 \\
GraphMAGE-SL ResGCN & 0.8002 & 0.7826 & 0.7781 & 0.7614 & 0.7612 & 0.7741 & 0.7860 & 0.8288 \\
Random Forest & 0.839 & 0.798 & 0.7844 & 0.7757 & 0.7659 & 0.7651 & 0.7753 & 0.7929 \\
Logistic Regression & 0.4843 & 0.4873 & 0.4724 & 0.4932 & 0.4726 & 0.519 & 0.5051 & 0.5026 \\
\midrule

\multicolumn{9}{c}{\textbf{F1 Score ($\uparrow$)}} \\
\midrule
Model & Week 0 & Week 1 & Week 2 & Week 3 & Week 4 & Week 5 & Week 6 & Week 7 \\
\midrule
GraphMAGE-SSL & \textbf{0.5867} & \textbf{0.5216} & 0.5106 & \textbf{0.4894} & \textbf{0.4745} & \textbf{0.4782} & \textbf{0.484} & 0.4949 \\
GraphMAGE-SL & 0.5848 & 0.5121 & \textbf{0.5121} & 0.4893 & 0.4719 & 0.4689 & 0.4736 & \textbf{0.4954} \\
MageNet Baseline & 0.4227 & 0.3929 & 0.3886 & 0.3763 & 0.3578 & 0.3733 & 0.3576 & 0.3671 \\
MageNet Baseline ResGCN & 0.3978 & 0.3598 & 0.3283 & 0.2816 & 0.2825 & 0.2411 & 0.3118 & 0.3301 \\
GraphMAGE-SL ResGCN & 0.4982 & 0.4696 & 0.4529 & 0.4254 & 0.4134 & 0.4137 & 0.4198 & 0.4571 \\
Random Forest & 0.3373 & 0.1857 & 0.0968 & 0.0343 & 0.0067 & 0.0001 & 0.0001 & 0 \\
Logistic Regression & 0.2832 & 0.2551 & 0.2417 & 0.2489 & 0.236 & 0.2505 & 0.2418 & 0.2266 \\
\midrule

\multicolumn{9}{c}{\textbf{Sensitivity ($\uparrow$)}} \\
\midrule
Model & Week 0 & Week 1 & Week 2 & Week 3 & Week 4 & Week 5 & Week 6 & Week 7 \\
\midrule
GraphMAGE-SSL & 0.7475 & 0.7 & 0.6703 & 0.6586 & 0.6036 & 0.6476 & 0.6602 & 0.704 \\
GraphMAGE-SL & 0.7483 & \textbf{0.7278} & 0.6739 & \textbf{0.6703} & 0.6066 & 0.6329 & 0.6767 & 0.6983 \\
MageNet Baseline & 0.619 & 0.5897 & 0.5836 & 0.5608 & 0.56 & 0.5913 & 0.5872 & 0.5942 \\
MageNet Baseline ResGCN & 0.5830 & 0.5475 & 0.5162 & 0.4300 & 0.4266 & 0.3803 & 0.5123 & 0.5448 \\
GraphMAGE-SL ResGCN & \textbf{0.7630} & 0.7124 & \textbf{0.7068} & 0.6646 & \textbf{0.6687} & \textbf{0.6777} & \textbf{0.7064} & \textbf{0.7900} \\
Random Forest & 0.215 & 0.1071 & 0.0525 & 0.0178 & 0.0034 & 0 & 0 & 0 \\
Logistic Regression & 0.4916 & 0.4442 & 0.4354 & 0.4389 & 0.4297 & 0.4414 & 0.4321 & 0.4007 \\
\bottomrule
\end{tabular}
}
\caption{Performance averaged over all nodes when the model has access to entire dataset.}
\label{tab:metrics_trainedOnAll_overall}
\end{table}

The random forest (RF) and logistic regression model (LR) both exhibit a common failure mode of classical approaches to WNV forecasting: low sensitivity. This severely limits their utility for longer range forecasting (5-8 weeks). In fact, the RF has zero sensitivity beyond week 4 meaning that all traps are predicted to be non-positive (no WNV). The LR model has a higher sensitivity (0.4-0.44) during this time period, but still well below the deep models (0.6-0.7). As indicated by the near 0.5 AUC, this is likely due the LR being no better than random guessing.

Each of the GNN based approaches exhibited a characteristic, ``U-shaped'', performance curve over time. Instead of F1, Sensitivity, and AUC decaying with prediction horizon, these metrics tended to rebound and improve starting at week five. As was shown in \cite{tonks2024forecasting}, this effect is likely due to the GNN learning to focus on trap information for near term prediction and environmental information for long term prediction. Our model shows a similar rebound pattern as the baseline GraphSAGE.

\subsection{Out-of-graph prediction}

Because new mosquito traps are introduced over time, and new districts are beginning to trap altogether, it is critical that our model can generalize to traps in unseen locations. In particular, we would like our model to generalize from high density urban areas, where a majority of trapping occurs, to rural areas which may lack the resource to run sophisticated mosquito surveillance programs. 

% Table \ref{tab:metrics_trainedOnAll_overall} show that our model generalizes 
% This is to test how well our model generalizes to new locations with different connectivity structure than the training data. For example, generalizing from rural to urban and urban to rural. 
To evaluate the performance of our models on unseen locations, we created two subsets of the data to use as training sets; denoted the \textit{Lower 80} and the \textit{Upper 80} respectively. 
% in handling scenarios where the training and test sets have different connectivity structures, we created two distinct data subsets: the Lower 80 Percent and the Upper 80 Percent. These subsets were designed based on the connectivity of nodes within the graph. A node’s connectivity was determined by summing the distances to its 10 nearest neighbors.
The Lower 80 subset includes all data except the 20\% most highly connected nodes, which were held out from the training set and reserved for testing. A node is deemed "highly connected" if the total weight of its edges to its nearest neighbors is higher than the 80th quantile of all node edge weight sums. By training on this subset we can evaluate how well our model generalizes to new nodes in highly connected regions, such as urban areas. 
Conversely, the Upper 80 subset includes all data except the 20\% least connected nodes were excluded from training and included in the test set. In this context, a node is categorized as "least connected" if the sum of its edge weights to its nearest neighbors is lower than 20th quantile of all node edge weight sums. This subset allows us to evaluate the model's ability to predict on sparsely connected or isolated nodes, i.e. rural areas, which are more challenging due to reduced contextual information from their neighbors and irregular monitoring patterns. 

\begin{table}[h]
\centering
\resizebox{\columnwidth}{!}{
\begin{tabular}{l|cccccccc}
\toprule

\multicolumn{9}{c}{\textbf{AUC Score ($\uparrow$)}} \\
\midrule
Model & Week 0 & Week 1 & Week 2 & Week 3 & Week 4 & Week 5 & Week 6 & Week 7 \\
\midrule

GraphMAGE-SSL & \textbf{0.8657} & \textbf{0.8341} & \textbf{0.8384} & \textbf{0.82} & 0.8147 & \textbf{0.8263} & \textbf{0.8437} & \textbf{0.8598} \\
GraphMAGE-SL & 0.8654 & 0.8338 & 0.8346 & 0.8193 & \textbf{0.8156} & 0.8195 & 0.8379 & 0.8592 \\
MageNet Baseline & 0.8237 & 0.789 & 0.7956 & 0.7807 & 0.782 & 0.7809 & 0.8054 & 0.8279 \\
MageNet Baseline ResGCN & 0.7581 & 0.7018 & 0.6683 & 0.6611 & 0.6595 & 0.6516 & 0.6738 & 0.6952 \\
GraphMAGE-SL ResGCN & 0.8019 & 0.7992 & 0.7874 & 0.7757 & 0.7697 & 0.7890 & 0.8066 & 0.8376 \\
Random Forest & 0.8406 & 0.8004 & 0.7837 & 0.767 & 0.756 & 0.7606 & 0.7843 & 0.8009 \\
Logistic Regression & 0.4775 & 0.4822 & 0.4633 & 0.4907 & 0.465 & 0.5238 & 0.5015 & 0.5002 \\
\midrule

\multicolumn{9}{c}{\textbf{F1 Score ($\uparrow$)}} \\
\midrule
Model & Week 0 & Week 1 & Week 2 & Week 3 & Week 4 & Week 5 & Week 6 & Week 7 \\
\midrule
GraphMAGE-SSL & 0.6365 & \textbf{0.5936} & \textbf{0.5692} & \textbf{0.5461} & 0.5225 & \textbf{0.5383} & \textbf{0.5507} & \textbf{0.5551} \\
GraphMAGE-SL & \textbf{0.6417} & 0.5882 & 0.5607 & 0.5457 & \textbf{0.525} & 0.5259 & 0.5353 & 0.5509 \\
MageNet Baseline & 0.5593 & 0.4549 & 0.4341 & 0.4251 & 0.3935 & 0.4091 & 0.4376 & 0.4265 \\
MageNet Baseline ResGCN & 0.4709 & 0.3588 & 0.2920 & 0.3039 & 0.2857 & 0.2885 & 0.2799 & 0.3057 \\
GraphMAGE-SL ResGCN & 0.5448 & 0.5524 & 0.5010 & 0.4886 & 0.4620 & 0.4732 & 0.5078 & 0.5406 \\
Random Forest & 0.3705 & 0.1976 & 0.0864 & 0.0087 & 0 & 0 & 0 & 0 \\
Logistic Regression & 0.2958 & 0.2808 & 0.253 & 0.2777 & 0.2521 & 0.2831 & 0.2679 & 0.2445 \\

\midrule

\multicolumn{9}{c}{\textbf{Sensitivity ($\uparrow$)}} \\
\midrule
Model & Week 0 & Week 1 & Week 2 & Week 3 & Week 4 & Week 5 & Week 6 & Week 7 \\
\midrule
GraphMAGE-SSL & \textbf{0.837} & 0.743 & 0.7518 & 0.7281 & 0.7179 & \textbf{0.764} & 0.8043 & 0.8558 \\
GraphMAGE-SL & 0.8275 & 0.7602 & \textbf{0.7744} & \textbf{0.742} & \textbf{0.7415} & 0.7434 & \textbf{0.8138} & 0.8558 \\
MageNet Baseline & 0.7181 & 0.5701 & 0.5425 & 0.5388 & 0.5154 & 0.5495 & 0.6441 & 0.6364 \\
MageNet Baseline ResGCN & 0.6154 & 0.4704 & 0.3877 & 0.4120 & 0.4034 & 0.3997 & 0.4329 & 0.4677 \\
GraphMAGE-SL ResGCN & 0.7796 & \textbf{0.7729} & 0.7283 & 0.6889 & 0.6730 & 0.7226 & 0.8101 & \textbf{0.8836} \\
Random Forest & 0.2412 & 0.1155 & 0.0464 & 0.0044 & 0 & 0 & 0 & 0 \\
Logistic Regression & 0.4201 & 0.3857 & 0.37 & 0.3945 & 0.3746 & 0.4 & 0.3903 & 0.3467 \\
\hline
\end{tabular}
}
\caption{Performance averaged over the 20 percent most connected nodes (Urban nodes) when the model is trained on the other 80 percent (Lower 80 percent).}
\label{tab:upper20_metrics_trainedOnLower80_weeks}
\end{table}

% Notes (Table 2)
% \begin{itemize}
%     \item Our method dominates F1 score meaning best classifier over entire interval. Improves over overall results because these points are easy to predict (compared to the average point). We are significantly better than others (including magenet). (1952 obs and 483 positives)
%     \item Possible rebounding (week 4)
%     \item mention Semisupervised didnt help that much overall using a nearest neighbor graph
%     \item reason/details: maintains high sensitivity and specificity over entire prediction interval
%     \item Random forest spec goes to 1, sense goes to 0 (collapses to all 0s)
%     \item Logistic regression does not collapse but is random guessing (AUC)
% \end{itemize}

Table \ref{tab:upper20_metrics_trainedOnLower80_weeks} shows the performance of all models trained on the Lower 80 subset and evaluated on the held-out upper 20\% most highly connected hold-out nodes. 
% At Week 0, these nodes accounted for 1952 observations, with 483 classified as positives. While the number of positives declines as the forecast horizon progresses, this reduction is not drastic.
% Our findings align with what our observations in Section \ref{sec:forecasting_skill}. 
As in Section \ref{sec:forecasting_skill}, the deep neural network (DNN) models consistently outperform classical models across all metrics. Notably, the random forest achieves a high AUC, but its sensitivity collapses to zero as it predicts exclusively negative outcomes, leading to specificity values approaching 1. Logistic regression, while avoiding collapse, performs as a random guesser, yielding AUC values that reflect poor discrimination between the classes. 
% This is likely due to the model simply not being able to hand the complexity of the data. 
The Mage Dense architecture consistently outperforms all other models, maintaining high sensitivity and specificity over the entire prediction interval. This robustness ensures strong classification performance, particularly for highly connected nodes, where accurate predictions are critical. Although semi-supervised Mage Dense offers only negligible improvements over its fully supervised counterpart, it, however, is still effective in forecasting.

%Lower 20

\begin{table}[h]
\centering
\resizebox{\columnwidth}{!}{
\begin{tabular}{l|cccccccc}
\toprule
\multicolumn{9}{c}{\textbf{AUC Score ($\uparrow$)}} \\
\midrule
Model & Week 0 & Week 1 & Week 2 & Week 3 & Week 4 & Week 5 & Week 6 & Week 7 \\
\midrule
GraphMAGE-SSL & 0.7621 & 0.7232 & 0.7008 & 0.6754 & 0.6879 & 0.694 & 0.716 & 0.7484 \\
GraphMAGE-SL & 0.7653 & 0.725 & 0.7079 & 0.6784 & 0.6924 & \textbf{0.6949} & \textbf{0.72} & \textbf{0.7497} \\
MageNet Baseline & 0.7148 & 0.6809 & 0.6683 & 0.6424 & 0.6406 & 0.6712 & 0.6755 & 0.7089 \\
MageNet Baseline ResGCN & 0.6584 & 0.6244 & 0.6188 & 0.5948 & 0.6052 & 0.6240 & 0.6426 & 0.6572 \\
GraphMAGE-SL ResGCN & 0.7219 & 0.6879 & 0.6757 & 0.6558 & 0.6681 & 0.6854 & 0.6994 & 0.7252 \\
Random Forest & \textbf{0.7801} & \textbf{0.7374} & \textbf{0.7214} & \textbf{0.6933} & \textbf{0.7026} & 0.6945 & 0.7137 & 0.7422 \\
Logistic Regression & 0.4953 & 0.4996 & 0.4815 & 0.5005 & 0.4959 & 0.509 & 0.5127 & 0.5012 \\
\midrule

\multicolumn{9}{c}{\textbf{F1 Score ($\uparrow$)}} \\
\midrule
GraphMAGE-SSL & 0.3438 & 0.287 & 0.2688 & 0.2468 & 0.2496 & 0.2421 & 0.2477 & 0.2511 \\
GraphMAGE-SL & \textbf{0.3451} & \textbf{0.2891} & \textbf{0.2781} & \textbf{0.2538} & \textbf{0.2499} & \textbf{0.2429} & \textbf{0.2475} & \textbf{0.2522} \\
MageNet Baseline & 0.266 & 0.2078 & 0.209 & 0.1874 & 0.1796 & 0.1871 & 0.1854 & 0.1793 \\
MageNet Baseline ResGCN & 0.2466 & 0.1834 & 0.1819 & 0.1628 & 0.1567 & 0.1638 & 0.1545 & 0.1549 \\
GraphMAGE-SL ResGCN & 0.3030 & 0.2523 & 0.2564 & 0.2307 & 0.2306 & 0.2298 & 0.2353 & 0.2166 \\
Random Forest & 0.1818 & 0.0454 & 0.0014 & 0 & 0 & 0 & 0 & 0 \\
Logistic Regression & 0.1961 & 0.1707 & 0.1621 & 0.1595 & 0.1586 & 0.1564 & 0.1531 & 0.1327 \\
\midrule

\multicolumn{9}{c}{\textbf{Sensitivity ($\uparrow$)}} \\
\midrule
GraphMAGE-SSL & \textbf{0.7603} & 0.612 & \textbf{0.5982} & \textbf{0.5667} & 0.5542 & 0.5669 & 0.6258 & \textbf{0.6808} \\
GraphMAGE-SL & 0.7599 & \textbf{0.6368} & 0.5804 & 0.5646 & \textbf{0.5801} & \textbf{0.5818} & \textbf{0.6319} & 0.6538 \\
MageNet Baseline & 0.6365 & 0.5579 & 0.5293 & 0.5117 & 0.4821 & 0.5128 & 0.5314 & 0.5009 \\
MageNet Baseline ResGCN & 0.5631 & 0.4407 & 0.4346 & 0.4117 & 0.3912 & 0.3885 & 0.3832 & 0.4476 \\
GraphMAGE-SL ResGCN & 0.6927 & 0.5512 & 0.5666 & 0.4769 & 0.5054 & 0.4995 & 0.5758 & 0.4988 \\
Random Forest & 0.1052 & 0.0238 & 0.0007 & 0 & 0 & 0 & 0 & 0 \\
Logistic Regression & 0.5917 & 0.5347 & 0.5171 & 0.516 & 0.5248 & 0.5149 & 0.5101 & 0.468 \\
\bottomrule
\end{tabular}
}
\caption{Performance averaged over the 20 least connected nodes (Rural nodes) when the model is trained on the other 80 percent (Upper 80 percent). We again observe metric rebound behavior, although it is less pronounced compared to the results in Table \ref{tab:metrics_trainedOnAll_overall}.}
\label{tab:Lower_20_metrics_trainedOnUpper80_weeks}
\end{table}

The results in Table \ref{tab:Lower_20_metrics_trainedOnUpper80_weeks} focus on the sparsely connected nodes that fall within the Lower 20 Percent of connectivity. These nodes, often located in more rural or isolated regions, are inherently challenging to classify due to their sparse connections, irregular checking patterns, and lower positivity rates. For the Lower 20 percent of nodes, Week 0 includes 2165 observations, of which 252 are positives. While the number of positives gradually declines over time, the overall ratio of positives to negatives remains relatively consistent.
Our proposed architecture, GraphMAGE, consistently achieves the highest F1 score across all forecast weeks, outperforming all other models. 

We note that the F1 scores in this subset are notably lower compared to the overall results presented in Table \ref{tab:metrics_trainedOnAll_overall}. This can seen in the substantial drop in sensitivity, reflecting the increased difficulty in identifying positives among sparsely connected nodes.
Interestingly, no rebound effect is observed in this evaluation, which contrasts with previous experiments.
% This absence may be attributed to the difficulty is modeling these rural and isolated nodes (Not sure what to say here so I just kinda said they are hard to classify).
The random forest model, as seen in prior experiments, collapses to predicting all negatives, achieving perfect specificity but zero sensitivity. Logistic regression does not collapse but instead exhibits random guessing behavior, as evidenced by its consistently low AUC.
While MageNet Baseline maintains better sensitivity compared to random forest and logistic regression, it fails to match the comprehensive performance of GraphMAGE. Notably, the semi-supervised version of GraphMAGE provides negligible improvement over the fully supervised implementation.

% \subsection{Something else?}

\subsection{Optimizing new trap placement}

Finally, as mosquito abatement programs grow, new traps need to be placed to improve surveillance. How to optimally place new traps so as to maximize their effectiveness is still an open question \citep{chakravarti2024place}. Taking inspiration from Bayesian optimization \citep{frazier2018tutorial, wang2023recent}, we propose to place traps in locations that maximally reduce overall predictive uncertainty. That is we will measure, how much traps in  different types of locations, when added to the model, reduces the overall predictive uncertainty. Due to data sharing restrictions, we break these results down by land use type, rather than specific latitude and longitude coordinates. Consequently, these findings may, therefore, be more applicable to regions beyond Illinois. 

% Because the proposed model can accurately predict to unseen locations, both urban (highly connected) and rural (sparsely connected), would like to use it to identify po

% % Idea: place traps in regions of high uncertainty (large loss). Need well calibrated uncertainty for this to work. Use isotonic regression to scale predictions. Identify regions by certain covariate categories.

% Accurately quantifying uncertainty is essential for assessing the reliability of our predictions, particularly in the context of varying node connectivity and across diverse environments. 

\paragraph{Isotonic probability calibration} 

Because our approach relies on the accuracy of the predicted probabilities, and deep neural networks are not well calibrated \citep{guo2017calibration}, we first apply isotonic calibration scheme \citep{berta2024classifier} to calibrate the predicted probabilities. Calibration ensures that the predicted probabilities are closer to the true probabilities, which makes our estimates of uncertainty more reliable.

% To enhance the reliability and interpretability of the model’s probabilistic predictions, we apply probability calibration using isotonic regression. Probability calibration ensures that the predicted probabilities reflect the true likelihood of an event occurring, (doing something like reduce over/under confidence). (citation). 

First, we train out model 20 times under 20 different seeds and average the predictions to obtain stable baseline predictive probability estimates. We then calibrate these probabilities using isotonic regression, which fits a non-decreasing real function to the predicted probabilities. This monotonic transformation preserves the order of predictions while correcting for miscalibrations. 
The final, calibrated, probability, \(\hat{P}\), is obtained using a weighted combination of the averaged predictions and the isotonic regression output:
\begin{equation}
\hat{P} = \lambda \bar{\hat{y}} + (1 - \lambda) f(\bar{\hat{y}})
\end{equation}
where \(\lambda\) is the weighting parameter, \(\bar{\hat{y}}\) is the mean prediction across 20 seeds, and \(f(\bar{\hat{y}})\) is the isotonic regression output \citep{berta2024classifier}. By fitting the isotonic regression on a separate calibration dataset, we prevent information leaking and preserve the integrity of the training and test sets.

% The isotonic regression solves the following optimization problem:
% \begin{equation}
% \min \sum_{i} w_i (y_i - \hat{y}_i)^2
% \end{equation}

% subject to the constraint:

% \begin{equation}
% \hat{y}_i \leq \hat{y}_j \quad \text{for } X_i \leq X_j
% \end{equation}

% where \(w_i\) are strictly positive weights, and \(y_i\) and \(\hat{y}_i\) represent the observed and predicted values, respectively. 

\paragraph{Uncertainty measure}
We use average entropy \citep{cover1999elements} of the (calibrated) predictions to determine which regions have high and low certainty. The average (over all prediction horizons) entropy at node $v$ is defined as
\begin{equation}
    H_v = \sum_{i = 1}^w \hat p_i \log \hat p_i + (1 - \hat p_i)\log(1-\hat p_i),
\end{equation}
where $w$ is the max forecasting lead time (8 weeks) and $\hat p_i$ is the predicted positive probability at node $v \in V$. Entropy serves as a measure of uncertainty, where higher entropy indicates less confident predictions. Entropy is calculated over multiple node subsets (All nodes, Upper 80, Lower 80), allowing us to evaluate how model confidence varies across regions with different levels of connectivity.

\paragraph{Uncertainty reduction by covariate} 

\begin{table}[ht!]
    \resizebox{\linewidth}{!}{%
    \centering
    \begin{tabular}{lcccccc}
        \toprule
          % & \multicolumn{6}{c}{\bf Trained on all nodes} \\
          \cmidrule(lr){2-7}
          & \multicolumn{3}{c}{\textbf{GraphMAGE-SSL}} & \multicolumn{3}{c}{\textbf{GraphMAGE-SL}} \\
          \cmidrule(lr){2-4} \cmidrule(lr){5-7}
          \textbf{Variable} & All Nodes & Upper 80 & Lower 80 & All Nodes & Upper 80 & Lower 80 \\
        \midrule
        \textit{Canopy}\\
    	\;\;Low & 0.3083 & 0.3592 & 0.3489 & 
                  0.3094 & 0.3619 & 0.3515 \\
        \;\;Medium & 0.3489 & 0.3879 & 0.3898 & 
                     0.3529 & 0.3925 & 0.3920 \\
        \;\;High & 0.2778 & 0.4027 & 0.3650 &
                   0.2873 & 0.4082 & 0.3624 \\

        \textit{Imperviousness}\\
        \;\;Low  & 0.3073 & 0.3635 & 0.3532 & 
                   0.3095 & 0.3657 & 0.3554 \\
        \;\;Medium  & 0.3406 & 0.3767 & 0.3717 & 
                      0.3407 & 0.3809 & 0.3750  \\
        \;\;High  & 0.3101 & 0.3767 & 0.3565 & 
                    0.3165 & 0.3809 & 0.3611  \\
        
        \textit{Land Usage}\\
        \;\;Open Water & 0.2736 & 0.3422 & 0.3270 & 
                         0.2779 & 0.3434 & 0.3301 \\
        \;\;Open Space Dev. & 0.3059 & 0.3697 & 0.3598 & 
                              0.3075 & 0.3716 & 0.3611 \\
        \;\;Low Dev. & 0.3298 & 0.3747 & 0.3688 &
                       0.3317 & 0.3774 & 0.3713 \\
        \;\;Medium Dev. & 0.3252 & 0.3704 & 0.3635 & 
                          0.3258 & 0.3809 & 0.3750 \\
        \;\;High Dev. & 0.3058 & 0.3673 & 0.3526 & 
                        0.3084 & 0.3713 & 0.3553 \\
        \;\;Deciduous Forest & 0.3240 & 0.3724 & 0.3597 & 
                               0.3248 & 0.3744 & 0.3599 \\
        \;\;Grasslands Herb. & 0.2793 & 0.3414 & 0.3249 & 
                               0.2800 & 0.3417 & 0.3243\\
        \;\;Pasture Hay & 0.2902 & 0.3712 & 0.3455 & 
                          0.2961 & 0.3720 & 0.3487 \\
        \;\;Cultivated Crops & 0.2668 & 0.3448 & 0.3192 & 
                               0.2684 & 0.3485 & 0.3232 \\
        \;\;Woody Wetlands & 0.3243 & 0.3646 & 0.3609 & 
                             0.3226 & 0.3681 & 0.3611 \\
        
        \textit{Roads}\\
        \;\;Primary Road & 0.2736 & 0.3422 & 0.3270 &
                           0.2779 & 0.3434 & 0.3301 \\
        \;\;Secondary Road & 0.3059 & 0.3697 & 0.3598 & 
                             0.3075 & 0.3716 & 0.3611\\
        \;\;Tertiary Road & 0.3298 & 0.3747 & 0.3688  & 
                            0.3317 & 0.3774 & 0.3713 \\
        \;\;Non Road & 0.3252 & 0.3704 & 0.3635 & 
                       0.3258 & 0.3734 & 0.3663 \\
        \bottomrule
      \end{tabular}
    }
    % \caption{Comparison (Entropy reduction?) of Mage Dense and Mage Dense Supervised models trained on all nodes and evaluated on the upper and lower 80\% of trap data.}
    \caption{Entropy by land usage type.}
    \label{tab:comparison_table_all}
\end{table}

The results, summarized in Table \ref{tab:comparison_table_all}, present the average entropy across three years of test data and averaged over the entire forecast horizon. Each column in the table represents a different training subset: All Nodes, Upper 80\% (urban-skewed), and Lower 80\% (rural-skewed) with all trained models evaluated on the same full test set. This design allows us to assess how training on different spatial subsets impacts uncertainty in predictions. 

The tables show several important findings. First, the semi-supervised Mage Dense architecture routinely outperforms its supervised counterpart. This is also reflected in its better performance on our metrics in section 4.2 and 4.3.  
Second, models trained on rural-skewed data (Lower 80) exhibit notably higher confidence, as reflected by consistently lower entropy across nearly all covariate classes. This suggests that rural training data provides greater diversity in environmental features and conditions, improving the model’s ability to generalize.
The model's uncertainty varies across different environmental variables. In regions with high tree canopy cover (prevalent in rural areas), the model achieves lower average entropy when trained on our rural proxy. Similarly, in areas with pasture hay, the model demonstrates lower uncertainty when trained on rural nodes as we would expect. The story is much the same for cultivated crops and other rural dominated variables. This advantage extends even to urban-dominated variables such as High Imperviousness, Open Space Developed, and High Developed. In all such cases, training on the rural proxy results in lower entropy than training on the urban proxy alone (see Table~\ref{tab:comparison_table_all}).

Training on rural nodes considerably reduces the overall entropy across the entire network. This effect is likely due to the increased diversity of environmental conditions present in the dataset, which enhances the model's adaptability to previously unseen regions. In contrast, training exclusively on urban nodes reduces generalization, leading to higher entropy when applied to rural areas and urban areas. The is easily visible by comparing the the Lower 80 and Upper 80 columns of table \ref{tab:comparison_table_all}. 

These results suggest practical implications for trap placement. The effective trapping of an area greatly depends on the optimality of the trap layout. Over concentration of trapping in an area could lead to lower predictive performance on nodes located elsewhere. Incorporating rural traps to ensure environmental diversity can help build more robust and confident predictive models.

\section{Discussion} 

% (summary of problem and proposed method)

% (summary of 4.2 and 4.3)

% (summary of 4.4) The uncertainty quantification results demonstrate the importance of incorporating diverse environmental conditions when training spatiotemporal models and how a diverse set of node locations can be of greater benefit than a single densely trapped area. Our findings reveal that including rural regions in the training set improves generalization, reducing entropy across the entire network. We also found that 

% (future work) mention limitations, estimate spatial graph. \\

West Nile virus is a present, and growing, threat to public health in the United States. Because there is no human vaccine, the best mitigation strategy is forecast when and where WNV may emerge each year. Current modeling strategies typically involve linear models with strong parametric assumptions, or simple machine learning approaches fit to highly aggregate features. Recent work on using Graph Neural Networks (GNNs) for WNV forecasting has shown that GNNs are a powerful and scalable alternative to traditional approaches. However, classic GNN models based on graph convolution suffer from degenercy issues with increasing depth, which limits their ability to process high resolution spatio-temporal data. 

We introduced a new Graph Neural Network (GNN) architecture, GraphMAGE, that uses a sparse skip connection strategy, to dramatically improve the stability and accuracy of large GNNs models. This new architecture allows us to train large GNN models and significantly out-perform baseline approaches a variety of WNV forecasting scenarios. In particular, we showed that our model consistently improves over the small GNN models introduced in \cite{tonks2024forecasting} and classical statistical approaches.
% In this work, we proposed the use of a Graph Neural Network based approach to accurately forecast West Nile virus prevalence as a proxy for a West Nile virus threat. This approach was motivated by the uneven distribution of mosquito traps and looking to maximize the use of spatial temporal weather and environmental data in Illinois. Our method utilizes spatial graphs built from trap locations and special case of ResGCN connections. This allows for better predictive accuracy and generalization including in areas that were previous unobserved. 

Our results show that our model successfully generalizes to both future realizations of nodes (mosquito traps) that were checked in the training set and to completely unseen nodes. We demonstrated that this result held in both highly connected (urban) regions and sparsely connected (rural) regions. This means that our model could potentially be adapted to new urban and rural locations, which have minimal to no data. 
% We evaluated the models performance across multiple experiments comparing other machine learning baselines to our model. The results show that our model significantly improved prediction accuracy over a longer forecast horizon including in regions with few traps. To compare to a situation where we would like to forecast in previously unobserved areas we created subsets of the entire data leaving the most connected nodes or the least connected nodes using these as proxies for urban and rural areas. We found that the GraphMAGE Architecture performs better in these unobserved regions than our comparative models.
We also demonstrated how our model could be used to suggest new trapping locations, We defined an optimal trap as one that maximally reduces uncertainty in the (calibrated) predictions. Table \ref{tab:comparison_table_all} showed that certain locations, such as high tree canopy regions, are highly uncertain in even the best models. Thus, by introducing new traps in these locations, we may be able to reduce overall uncertainty more efficiently than if we had introduced them in regions of already high certainty, such as low to medium developed areas.

% The uncertainty quantification results demonstrate the importance of incorporating diverse environmental conditions when training spatiotemporal models and how a diverse set of node locations can be of greater benefit than a single densely trapped area. Our findings reveal that including rural regions in the training set improves generalization, reducing entropy across the entire network. We also found that 

Although our model improves over existing works, there are still remain several limitations. The first is that our approach relies on pre-specified graph. Dynamically learning graphs as part of the model may be able to improve performance. This may also allow us to incorporate covariates into the connection scheme, rather than relying purely on spatial distance. Second, our approach still uses some data pre-processing to improve performance. More general approaches, that can learn from purely raw data may be able to improve even further. However, this may require significantly more data than is currently available. 

\section*{Acknowledgments}

This research was generated from the Midwest Center of Excellence in Vector-Borne Disease which is funded by the cooperative agreement U01CK000505 from the Centers for Disease Control and Prevention. The contents are solely the responsibility of the authors and do not necessarily represent the official views of the Centers for Disease Control and Prevention. \\

\noindent Portions of this research were conducted with the advanced computing resources
provided by Texas A\&M Department of Statistics Arseven Computing Cluster.

% Several limitations suggest interesting future suggestions: the high dimensional data combined with the message passing can be slow to train especially in the non-JAX environment, training stability, hyperparameter tuning, data privacy issues hindered exploration of various connection strategies, and ...

% \subsubsection*{Author Contributions}
% If you'd like to, you may include  a section for author contributions as is done
% in many journals. This is optional and at the discretion of the authors.

% \subsubsection*{Acknowledgments}
% Use unnumbered third level headings for the acknowledgments. All
% acknowledgments, including those to funding agencies, go at the end of the paper.

\bibliography{main_th}
\bibliographystyle{apalike} 

\appendix
\section{Appendix}

\subsection{Implementation and training details} \label{adx:sec:implementation}

% In particular, we will use environmental identifiers, gradient scheduling, adaptive gradient clipping, spatial weighted graphs, and a semi-supervised learning framework. 

An $L$-layer graph mage is an efficient GraphSAGE ensemble consisting of a single $1$-Layer, $2$-layer, ..., and $L$-layer GraphSAGE model. However, training the ``deep paths'' of a deep GraphMAGE is still plagued with instabilities from lack of residuals leading to backprop issues and overly smoothing from neighbor aggregation. Furthermore, our effective ensemble is exponentially smaller than a full resent, so we may see less benefit in terms of loss smoothing or preventing gradient shattering.

%We use the following techniques to improve the overall performance:  gradient scheduling, adaptive gradient clipping, and semi-supervised training.

%comment: We set max neighbors to 10 and performed sensitivity analysis that showed performance did not change by adding more neighbors. Something about information saturation.

To minimize training instabilities, we implemented: early stopping, cosine decay learning rate scheduling, and adaptive gradient clipping (AGC) \citep{brock2021high}. Among these strategies, AGC helped the most in encouraging stable and efficient convergence. Gradient clipping techniques have been shown to stabilize training, accelerate convergence, and enable the use of larger learning rates \citep{zhang2019gradient}. The adaptive gradient clipping approach introduced by \citet{brock2021high} operates by scaling gradients on a parameter basis, ensuring that the gradient norm does not exceed a threshold relative to the norm of the parameter itself. This is a more scale aware approach compared to traditional gradient clipping and helps prevent gradient explosion.

\subsection{Trained on all - eval on partition}

\begin{table}[!ht]
\scriptsize
\centering
\resizebox{\columnwidth}{!}{
\begin{tabular}{l|cccccccc}
\toprule
\multicolumn{9}{c}{\textbf{AUC Score ($\uparrow$)}} \\
\midrule
Model & Week 0 & Week 1 & Week 2 & Week 3 & Week 4 & Week 5 & Week 6 & Week 7 \\
\midrule
GraphMAGE-SSL & 0.8516 & 0.8177 & 0.8089 & 0.8047 & 0.8026 & 0.813 & 0.8234 & 0.845 \\
GraphMAGE-SL & 0.8508 & 0.8164 & 0.8089 & 0.8018 & 0.7963 & 0.8056 & 0.8156 & 0.8443 \\
MageNet Baseline & 0.8059 & 0.7812 & 0.781 & 0.7747 & 0.7782 & 0.7828 & 0.7954 & 0.8147 \\
MageNet Baseline ResGCN & 0.7322 & 0.7074 & 0.6977 & 0.6664 & 0.6950 & 0.6693 & 0.7180 & 0.7379 \\
GraphMAGE-SL ResGCN & 0.8002 & 0.7826 & 0.7781 & 0.7614 & 0.7612 & 0.7741 & 0.7860 & 0.8288 \\
Random Forest & 0.839 & 0.798 & 0.7844 & 0.7757 & 0.7659 & 0.7651 & 0.7753 & 0.7929 \\
Logistic Regression & 0.4843 & 0.4873 & 0.4724 & 0.4932 & 0.4726 & 0.519 & 0.5051 & 0.5026 \\
\midrule

\multicolumn{9}{c}{\textbf{Accuracy ($\uparrow$)}} \\
\midrule
Model & Week 0 & Week 1 & Week 2 & Week 3 & Week 4 & Week 5 & Week 6 & Week 7 \\
\midrule
GraphMAGE-SSL & 0.7844 & 0.7626 & 0.7703 & 0.7596 & 0.7702 & 0.7664 & 0.7717 & 0.7773 \\
GraphMAGE-SL & 0.7827 & 0.7437 & 0.7706 & 0.7551 & 0.7666 & 0.7629 & 0.7557 & 0.7799 \\
MageNet Baseline & 0.7269 & 0.7278 & 0.7326 & 0.7339 & 0.7226 & 0.7248 & 0.7307 & 0.7486 \\
MageNet Baseline ResGCN & 0.7250 & 0.7240 & 0.7096 & 0.7411 & 0.7490 & 0.7629 & 0.7419 & 0.7584 \\
GraphMAGE-SL ResGCN & 0.7060 & 0.7208 & 0.7130 & 0.7229 & 0.7108 & 0.7193 & 0.7209 & 0.7270 \\
Random Forest & 0.8274 & 0.8268 & 0.8253 & 0.825 & 0.8275 & 0.8346 & 0.8376 & 0.8451 \\
Logistic Regression & 0.4906 & 0.5194 & 0.5103 & 0.5329 & 0.5176 & 0.5636 & 0.5582 & 0.5755 \\
\midrule

\multicolumn{9}{c}{\textbf{Brier Score ($\downarrow$)}} \\
\midrule
Model & Week 0 & Week 1 & Week 2 & Week 3 & Week 4 & Week 5 & Week 6 & Week 7 \\
\midrule
GraphMAGE-SSL & 0.1467 & 0.1576 & 0.1544 & 0.1588 & 0.1517 & 0.1541 & 0.1485 & 0.1415 \\
GraphMAGE-SL & 0.1462 & 0.1669 & 0.1536 & 0.1624 & 0.154 & 0.1543 & 0.1571 & 0.14 \\
MageNet Baseline & 0.1842 & 0.1878 & 0.1824 & 0.1807 & 0.1828 & 0.1749 & 0.1729 & 0.1622 \\
MageNet Baseline ResGCN & 0.2114 & 0.2111 & 0.2248 & 0.1960 & 0.1847 & 0.1749 & 0.1748 & 0.1630 \\
GraphMAGE-SL ResGCN & 0.1874 & 0.1802 & 0.1837 & 0.1825 & 0.1876 & 0.1806 & 0.1707 & 0.1632 \\
Random Forest & 0.1216 & 0.1237 & 0.1258 & 0.1281 & 0.129 & 0.1266 & 0.1239 & 0.117 \\
Logistic Regression & 0.3095 & 0.3011 & 0.3075 & 0.2937 & 0.3023 & 0.2671 & 0.2782 & 0.2611 \\
\midrule

\multicolumn{9}{c}{\textbf{F1 Score ($\uparrow$)}} \\
\midrule
Model & Week 0 & Week 1 & Week 2 & Week 3 & Week 4 & Week 5 & Week 6 & Week 7 \\
\midrule
GraphMAGE-SSL & 0.5867 & 0.5216 & 0.5106 & 0.4894 & 0.4745 & 0.4782 & 0.484 & 0.4949 \\
GraphMAGE-SL & 0.5848 & 0.5121 & 0.5121 & 0.4893 & 0.4719 & 0.4689 & 0.4736 & 0.4954 \\
MageNet Baseline & 0.4227 & 0.3929 & 0.3886 & 0.3763 & 0.3578 & 0.3733 & 0.3576 & 0.3671 \\
MageNet Baseline ResGCN & 0.3978 & 0.3598 & 0.3283 & 0.2816 & 0.2825 & 0.2411 & 0.3118 & 0.3301 \\
GraphMAGE-SL ResGCN & 0.4982 & 0.4696 & 0.4529 & 0.4254 & 0.4134 & 0.4137 & 0.4198 & 0.4571 \\
Random Forest & 0.3373 & 0.1857 & 0.0968 & 0.0343 & 0.0067 & 0.0001 & 0.0001 & 0 \\
Logistic Regression & 0.2832 & 0.2551 & 0.2417 & 0.2489 & 0.236 & 0.2505 & 0.2418 & 0.2266 \\
\midrule

\multicolumn{9}{c}{\textbf{Specificity ($\uparrow$)}} \\
\midrule
Model & Week 0 & Week 1 & Week 2 & Week 3 & Week 4 & Week 5 & Week 6 & Week 7 \\
\midrule
GraphMAGE-SSL & 0.7938 & 0.7768 & 0.792 & 0.7809 & 0.8049 & 0.7899 & 0.7933 & 0.7907 \\
GraphMAGE-SL & 0.7915 & 0.7473 & 0.7916 & 0.773 & 0.7999 & 0.7887 & 0.771 & 0.7949 \\
MageNet Baseline & 0.7546 & 0.759 & 0.7649 & 0.7705 & 0.7565 & 0.7513 & 0.7585 & 0.7768 \\
MageNetBaseline ResGCN & 0.7615 & 0.7640 & 0.7516 & 0.8069 & 0.8160 & 0.8387 & 0.7864 & 0.7975 \\
GraphMAGE-SL ResGCN & 0.6914 & 0.7227 & 0.7144 & 0.7352 & 0.7195 & 0.7276 & 0.7237 & 0.7154 \\
Random Forest & 0.9846 & 0.9894 & 0.9931 & 0.9959 & 0.9989 & 0.9999 & 1 & 1 \\
Logistic Regression & 0.4904 & 0.5364 & 0.5266 & 0.5528 & 0.5359 & 0.5878 & 0.5826 & 0.6076 \\
\midrule

\multicolumn{9}{c}{\textbf{Sensitivity ($\uparrow$)}} \\
\midrule
Model & Week 0 & Week 1 & Week 2 & Week 3 & Week 4 & Week 5 & Week 6 & Week 7 \\
\midrule
GraphMAGE-SSL & 0.7475 & 0.7 & 0.6703 & 0.6586 & 0.6036 & 0.6476 & 0.6602 & 0.704 \\
GraphMAGE-SL & 0.7483 & 0.7278 & 0.6739 & 0.6703 & 0.6066 & 0.6329 & 0.6767 & 0.6983 \\
MageNet Baseline & 0.619 & 0.5897 & 0.5836 & 0.5608 & 0.56 & 0.5913 & 0.5872 & 0.5942 \\
MageNet Baseline ResGCN & 0.5830 & 0.5475 & 0.5162 & 0.4300 & 0.4266 & 0.3803 & 0.5123 & 0.5448 \\
GraphMAGE-SL ResGCN & 0.7630 & 0.7124 & 0.7068 & 0.6646 & 0.6687 & 0.6777 & 0.7064 & 0.7900 \\
Random Forest & 0.215 & 0.1071 & 0.0525 & 0.0178 & 0.0034 & 0 & 0 & 0 \\
Logistic Regression & 0.4916 & 0.4442 & 0.4354 & 0.4389 & 0.4297 & 0.4414 & 0.4321 & 0.4007 \\
\bottomrule
\end{tabular}
}
\caption{Model evaluation metrics when trained on all nodes over 8 weeks}
\label{tab:metrics_trainedOnAll_overall_appendix}
\end{table}

%upper 20 trained on all table
\begin{table}[!ht]
\scriptsize
\centering
\resizebox{\columnwidth}{!}{
\begin{tabular}{l|cccccccc}

\toprule
\multicolumn{9}{c}{\textbf{AUC Score ($\uparrow$)}} \\
\midrule
Model & Week 0 & Week 1 & Week 2 & Week 3 & Week 4 & Week 5 & Week 6 & Week 7 \\
\midrule
GraphMAGE-SSL & 0.8698 & 0.8353 & 0.8305 & 0.8151 & 0.8053 & 0.8101 & 0.8331 & 0.8494 \\
GraphMAGE-SL & 0.8685 & 0.8309 & 0.8303 & 0.8073 & 0.7949 & 0.7994 & 0.8174 & 0.8469 \\
MageNet Baseline & 0.8155 & 0.7859 & 0.7953 & 0.7783 & 0.7799 & 0.7838 & 0.8017 & 0.8211 \\
MageNet Baseline ResGCN & 0.7565 & 0.7253 & 0.7177 & 0.6683 & 0.7019 & 0.6796 & 0.7275 & 0.7496 \\
GraphMAGE-SL ResGCN & 0.8100 & 0.7897 & 0.7882 & 0.7683 & 0.7694 & 0.7806 & 0.7994 & 0.8298 \\
Random Forest & 0.846 & 0.8044 & 0.7922 & 0.7745 & 0.7642 & 0.7664 & 0.786 & 0.8026 \\
Logistic Regression & 0.4805 & 0.4835 & 0.4638 & 0.4904 & 0.4645 & 0.5245 & 0.5023 & 0.5018 \\
\midrule

\multicolumn{9}{c}{\textbf{Accuracy ($\uparrow$)}} \\
\midrule
Model & Week 0 & Week 1 & Week 2 & Week 3 & Week 4 & Week 5 & Week 6 & Week 7  \\
\midrule
GraphMAGE-SSL & 0.7861 & 0.7524 & 0.7564 & 0.7371 & 0.7363 & 0.7325 & 0.7572 & 0.7634 \\
GraphMAGE-SL & 0.7876 & 0.7311 & 0.7621 & 0.7277 & 0.729 & 0.7266 & 0.7342 & 0.7636 \\
MageNet Baseline & 0.7226 & 0.72 & 0.7262 & 0.7176 & 0.7038 & 0.7065 & 0.7177 & 0.739 \\
MageNet Baseline ResGCN & 0.7236 & 0.7146 & 0.7001 & 0.7197 & 0.7287 & 0.7429 & 0.7306 & 0.7507 \\
GraphMAGE-SL ResGCN & 0.7013 & 0.7164 & 0.7008 & 0.7027 & 0.6866 & 0.6902 & 0.7041 & 0.7148 \\
Random Forest & 0.8018 & 0.7863 & 0.7921 & 0.7849 & 0.7928 & 0.7992 & 0.8038 & 0.813 \\
Logistic Regression & 0.4881 & 0.5156 & 0.5037 & 0.5284 & 0.5099 & 0.5629 & 0.5517 & 0.5687 \\
\midrule

\multicolumn{9}{c}{\textbf{Brier Score ($\downarrow$)}} \\
\midrule
Model & Week 0 & Week 1 & Week 2 & Week 3 & Week 4 & Week 5 & Week 6 & Week 7  \\
\midrule
GraphMAGE-SSL & 0.1526 & 0.1681 & 0.1662 & 0.1725 & 0.1696 & 0.1742 & 0.1602 & 0.1528 \\
GraphMAGE-SL & 0.1502 & 0.1794 & 0.162 & 0.1788 & 0.1734 & 0.1758 & 0.1731 & 0.152 \\
MageNet Baseline & 0.1846 & 0.1916 & 0.1843 & 0.1866 & 0.1891 & 0.1811 & 0.1773 & 0.1647 \\
MageNet Baseline ResGCN & 0.2141 & 0.2215 & 0.2340 & 0.2137 & 0.1985 & 0.1905 & 0.1827 & 0.1709 \\
GraphMAGE-SL ResGCN & 0.1954 & 0.1900 & 0.1949 & 0.1994 & 0.2041 & 0.1978 & 0.1826 & 0.1753 \\
Random Forest & 0.137 & 0.1463 & 0.1456 & 0.1527 & 0.1512 & 0.1503 & 0.1456 & 0.1374 \\
Logistic Regression & 0.3101 & 0.3043 & 0.3109 & 0.2973 & 0.307 & 0.267 & 0.2829 & 0.2638 \\
\midrule

\multicolumn{9}{c}{\textbf{F1 Score ($\uparrow$)}} \\
\midrule
Model & Week 0 & Week 1 & Week 2 & Week 3 & Week 4 & Week 5 & Week 6 & Week 7 \\
\midrule
GraphMAGE-SSL & 0.6557 & 0.5914 & 0.5704 & 0.5379 & 0.5064 & 0.5205 & 0.552 & 0.5581 \\
GraphMAGE-SL & 0.655 & 0.5762 & 0.5742 & 0.5304 & 0.4948 & 0.5074 & 0.5286 & 0.5515 \\
MageNet Baseline & 0.4687 & 0.4447 & 0.436 & 0.4183 & 0.3884 & 0.4138 & 0.3993 & 0.409 \\
MageNet Baseline ResGCN & 0.4473 & 0.4109 & 0.3683 & 0.3130 & 0.3083 & 0.2695 & 0.3510 & 0.3757 \\
GraphMAGE-SL ResGCN & 0.5504 & 0.5291 & 0.4993 & 0.4662 & 0.4458 & 0.4474 & 0.4701 & 0.5102 \\
Random Forest & 0.4078 & 0.2302 & 0.121 & 0.0356 & 0.003 & 0 & 0 & 0 \\
Logistic Regression & 0.317 & 0.2954 & 0.2677 & 0.2865 & 0.261 & 0.2926 & 0.2744 & 0.2557 \\
\midrule

\multicolumn{9}{c}{\textbf{Specificity ($\uparrow$)}} \\
\midrule
Model & Week 0 & Week 1 & Week 2 & Week 3 & Week 4 & Week 5 & Week 6 & Week 7\\
\midrule
GraphMAGE-SSL & 0.7744 & 0.7466 & 0.757 & 0.7436 & 0.7564 & 0.7347 & 0.7557 & 0.7554 \\
GraphMAGE-SL & 0.7788 & 0.714 & 0.7659 & 0.7308 & 0.7505 & 0.7329 & 0.7284 & 0.76 \\
MageNet Baseline & 0.7493 & 0.7533 & 0.7581 & 0.7543 & 0.7367 & 0.728 & 0.7418 & 0.7662 \\
MageNet Baseline ResGCN & 0.7619 & 0.7577 & 0.7433 & 0.7948 & 0.8033 & 0.8300 & 0.7783 & 0.7899 \\
GraphMAGE-SL ResGCN & 0.6709 & 0.7107 & 0.6893 & 0.7057 & 0.6837 & 0.6845 & 0.6882 & 0.6831 \\
Random Forest & 0.9747 & 0.9816 & 0.9897 & 0.9937 & 0.998 & 1 & 1 & 1 \\
Logistic Regression & 0.4904 & 0.5389 & 0.5263 & 0.5532 & 0.5338 & 0.5911 & 0.5816 & 0.6086 \\
\midrule

\multicolumn{9}{c}{\textbf{Sensitivity ($\uparrow$)}} \\
\midrule
Model & Week 0 & Week 1 & Week 2 & Week 3 & Week 4 & Week 5 & Week 6 & Week 7  \\
\midrule
GraphMAGE-SSL & 0.8216 & 0.7718 & 0.7542 & 0.7132 & 0.6588 & 0.7236 & 0.7631 & 0.7984 \\
GraphMAGE-SL & 0.8142 & 0.7879 & 0.7482 & 0.7163 & 0.6463 & 0.7014 & 0.7584 & 0.7793 \\
MageNet Baseline & 0.6415 & 0.6093 & 0.6093 & 0.583 & 0.577 & 0.6209 & 0.6192 & 0.6205 \\
MageNet Baseline ResGCN & 0.6071 & 0.5715 & 0.5415 & 0.4445 & 0.4409 & 0.3963 & 0.5352 & 0.5800 \\
GraphMAGE-SL ResGCN & 0.7938 & 0.7355 & 0.7428 & 0.6916 & 0.6980 & 0.7128 & 0.7695 & 0.8526 \\
Random Forest & 0.2758 & 0.1381 & 0.0669 & 0.0185 & 0.0015 & 0 & 0 & 0 \\
Logistic Regression & 0.4813 & 0.4383 & 0.4208 & 0.4374 & 0.4177 & 0.4509 & 0.4292 & 0.3955 \\
\bottomrule
\end{tabular}
}
\caption{Model evaluation metrics for the 20 percent most connected nodes when trained on all nodes over 8 weeks}
\label{tab:metrics_trainedOnAll_upper20_appendix}
\end{table}

%Lower 20 trained on all table
\begin{table}[!ht]
\scriptsize
\centering
\resizebox{\columnwidth}{!}{
\begin{tabular}{l|cccccccc}

\toprule
\multicolumn{8}{c}{\textbf{AUC Score ($\uparrow$)}} \\
\midrule
Model & Week 0 & Week 1 & Week 2 & Week 3 & Week 4 & Week 5 & Week 6 & Week 7 \\
\midrule
GraphMAGE-SSL & 0.7555 & 0.7193 & 0.7051 & 0.676 & 0.704 & 0.7193 & 0.722 & 0.751 \\
GraphMAGE-SL & 0.7526 & 0.7214 & 0.7086 & 0.678 & 0.7028 & 0.713 & 0.7189 & 0.7475 \\
MageNet Baseline & 0.7138 & 0.6833 & 0.6788 & 0.6486 & 0.6654 & 0.6822 & 0.7048 & 0.7344 \\
MageNet Baseline ResGCN & 0.6573 & 0.6273 & 0.6193 & 0.5990 & 0.6211 & 0.6076 & 0.6568 & 0.6768 \\
GraphMAGE-SL ResGCN & 0.6954 & 0.6734 & 0.6610 & 0.6407 & 0.6458 & 0.6670 & 0.6806 & 0.7220 \\
Random Forest & 0.7847 & 0.7368 & 0.7218 & 0.6976 & 0.7061 & 0.6942 & 0.7138 & 0.7401 \\
Logistic Regression & 0.494 & 0.4978 & 0.4809 & 0.5003 & 0.4964 & 0.5097 & 0.5128 & 0.5011 \\
\midrule

\multicolumn{8}{c}{\textbf{Accuracy ($\uparrow$)}} \\
\midrule
Model & Week 0 & Week 1 & Week 2 & Week 3 & Week 4 & Week 5 & Week 6 & Week 7 \\
\midrule
GraphMAGE-SSL & 0.7942 & 0.8129 & 0.8325 & 0.8096 & 0.8365 & 0.833 & 0.8254 & 0.8316 \\
GraphMAGE-SL & 0.7845 & 0.787 & 0.8212 & 0.8013 & 0.8268 & 0.8181 & 0.8099 & 0.8266 \\
MageNet Baseline & 0.747 & 0.7656 & 0.7783 & 0.7734 & 0.7699 & 0.7748 & 0.7788 & 0.7897 \\
MageNet Baseline ResGCN & 0.7359 & 0.7505 & 0.7505 & 0.7949 & 0.7943 & 0.8092 & 0.7753 & 0.7938 \\
GraphMAGE-SL ResGCN & 0.7124 & 0.7466 & 0.7522 & 0.7704 & 0.7537 & 0.7671 & 0.7623 & 0.7649 \\
Random Forest & 0.889 & 0.8998 & 0.8993 & 0.9048 & 0.9058 & 0.9099 & 0.9132 & 0.9219 \\
Logistic Regression & 0.4891 & 0.5333 & 0.5218 & 0.5427 & 0.5349 & 0.5697 & 0.5703 & 0.5902 \\
\midrule

\multicolumn{8}{c}{\textbf{Brier Score ($\downarrow$)}} \\
\midrule
Model & Week 0 & Week 1 & Week 2 & Week 3 & Week 4 & Week 5 & Week 6 & Week 7 \\
\midrule
GraphMAGE-SSL & 0.1327 & 0.127 & 0.119 & 0.1301 & 0.1162 & 0.1158 & 0.1176 & 0.1115 \\
GraphMAGE-SL & 0.1368 & 0.1395 & 0.1244 & 0.1343 & 0.1202 & 0.1205 & 0.1249 & 0.1133 \\
MageNet Baseline & 0.1754 & 0.1704 & 0.1639 & 0.1646 & 0.1635 & 0.1567 & 0.1564 & 0.1489 \\
MageNet Baseline ResGCN & 0.2008 & 0.1889 & 0.1940 & 0.1529 & 0.1510 & 0.1385 & 0.1497 & 0.1356 \\
GraphMAGE-SL ResGCN & 0.1714 & 0.1574 & 0.1565 & 0.1494 & 0.1561 & 0.1494 & 0.1435 & 0.1376 \\
Random Forest & 0.088 & 0.0828 & 0.0842 & 0.083 & 0.0827 & 0.0806 & 0.0773 & 0.0703 \\
Logistic Regression & 0.3075 & 0.2922 & 0.299 & 0.285 & 0.2896 & 0.2635 & 0.2698 & 0.2522 \\
\midrule

\multicolumn{8}{c}{\textbf{F1 Score ($\uparrow$)}} \\
\midrule
Model & Week 0 & Week 1 & Week 2 & Week 3 & Week 4 & Week 5 & Week 6 & Week 7 \\
\midrule
GraphMAGE-SSL & 0.3426 & 0.274 & 0.2278 & 0.2126 & 0.1936 & 0.1871 & 0.1829 & 0.1863 \\
GraphMAGE-SL & 0.3398 & 0.2914 & 0.2526 & 0.2281 & 0.2132 & 0.1945 & 0.2056 & 0.1915 \\
MageNet Baseline & 0.2581 & 0.2188 & 0.221 & 0.1812 & 0.1828 & 0.1841 & 0.1917 & 0.1913 \\
MageNet Baseline ResGCN & 0.2423 & 0.1923 & 0.1748 & 0.1439 & 0.1432 & 0.1196 & 0.1641 & 0.1574 \\
GraphMAGE-SL ResGCN & 0.2904 & 0.2540 & 0.2318 & 0.1928 & 0.1821 & 0.1836 & 0.1858 & 0.2090 \\
Random Forest & 0.1441 & 0.0203 & 0 & 0 & 0 & 0 & 0 & 0 \\
Logistic Regression & 0.1908 & 0.1635 & 0.1564 & 0.1532 & 0.1546 & 0.153 & 0.1518 & 0.1306 \\
\midrule

\multicolumn{8}{c}{\textbf{Specificity ($\uparrow$)}} \\
\midrule
Model & Week 0 & Week 1 & Week 2 & Week 3 & Week 4 & Week 5 & Week 6 & Week 7 \\
\midrule
GraphMAGE-SSL & 0.8378 & 0.8631 & 0.8969 & 0.8656 & 0.9013 & 0.8935 & 0.8817 & 0.8802 \\
GraphMAGE-SL & 0.8252 & 0.8255 & 0.8784 & 0.8523 & 0.886 & 0.8746 & 0.8596 & 0.8739 \\
MageNet Baseline & 0.7869 & 0.8077 & 0.8241 & 0.8228 & 0.8159 & 0.8186 & 0.8183 & 0.8241 \\
MageNet Baseline ResGCN & 0.7742 & 0.7929 & 0.7980 & 0.8516 & 0.8503 & 0.8667 & 0.8166 & 0.8332 \\
GraphMAGE-SL ResGCN & 0.7359 & 0.7789 & 0.7920 & 0.8177 & 0.7979 & 0.8108 & 0.8007 & 0.7935 \\
Random Forest & 0.9955 & 0.9984 & 0.9994 & 1 & 1 & 1 & 1 & 1 \\
Logistic Regression & 0.4856 & 0.542 & 0.5304 & 0.554 & 0.5434 & 0.5832 & 0.5821 & 0.6065 \\
\midrule

\multicolumn{8}{c}{\textbf{Sensitivity ($\uparrow$)}} \\
\midrule
Model & Week 0 & Week 1 & Week 2 & Week 3 & Week 4 & Week 5 & Week 6 & Week 7 \\
\midrule
GraphMAGE-SSL & 0.4631 & 0.3593 & 0.2539 & 0.2769 & 0.2132 & 0.2218 & 0.2332 & 0.2577 \\
GraphMAGE-SL & 0.4758 & 0.4398 & 0.3083 & 0.316 & 0.2574 & 0.2472 & 0.2872 & 0.2683 \\
MageNet Baseline & 0.4442 & 0.3856 & 0.3673 & 0.3034 & 0.3275 & 0.3315 & 0.3628 & 0.3834 \\
MageNet Baseline ResGCN  & 0.4454 & 0.3676 & 0.3240 & 0.2556 & 0.2561 & 0.2280 & 0.3415 & 0.3281 \\
GraphMAGE-SL ResGCN & 0.5339 & 0.4556 & 0.3956 & 0.3206 & 0.3294 & 0.3259 & 0.3585 & 0.4272 \\
Random Forest & 0.0804 & 0.0104 & 0 & 0 & 0 & 0 & 0 & 0 \\
Logistic Regression & 0.5155 & 0.4551 & 0.444 & 0.4354 & 0.4529 & 0.4331 & 0.4465 & 0.397 \\
\bottomrule
\end{tabular}
}
\caption{Model evaluation metrics for the 20 percent least connected nodes when trained on all nodes over 8 weeks}
\label{tab:metrics_trainedOnAll_lower20_appendix}
\end{table}

%Lower 20 percent held out 

%Overall 
\begin{table}[ht!]
\scriptsize
\centering
\resizebox{\columnwidth}{!}{
\begin{tabular}{l|cccccccc}
\toprule

\multicolumn{9}{c}{\textbf{AUC Score ($\uparrow$)}} \\
\midrule
Model & Week 0 & Week 1 & Week 2 & Week 3 & Week 4 & Week 5 & Week 6 & Week 7 \\
\midrule

GraphMAGE-SSL & 0.8464 & 0.8152 & 0.8096 & 0.811 & 0.8071 & 0.8142 & 0.8222 & 0.8454 \\
GraphMAGE-SL & 0.8472 & 0.8151 & 0.8105 & 0.8113 & 0.8069 & 0.8124 & 0.8218 & 0.8447 \\
MageNet Baseline & 0.7948 & 0.7712 & 0.7691 & 0.771 & 0.7604 & 0.7777 & 0.765 & 0.7793 \\
MageNet Baseline ResGCN & 0.7287 & 0.6966 & 0.7010 & 0.6790 & 0.6758 & 0.7141 & 0.7080 & 0.7210 \\
GraphMAGE-SL ResGCN & 0.8045 & 0.7687 & 0.7816 & 0.7839 & 0.7828 & 0.7912 & 0.7994 & 0.8201 \\
Random Forest & 0.8381 & 0.7985 & 0.785 & 0.7745 & 0.7638 & 0.7663 & 0.7735 & 0.7928 \\
Logistic Regression & 0.4863 & 0.4892 & 0.4723 & 0.4923 & 0.4708 & 0.522 & 0.5061 & 0.5037 \\
\midrule
\multicolumn{9}{c}{\textbf{Accuracy ($\uparrow$)}} \\
\midrule
Model & Week 0 & Week 1 & Week 2 & Week 3 & Week 4 & Week 5 & Week 6 & Week 7 \\
\midrule
GraphMAGE-SSL & 0.7391 & 0.7289 & 0.7077 & 0.7022 & 0.707 & 0.7083 & 0.696 & 0.7207 \\
GraphMAGE-SL & 0.7383 & 0.7177 & 0.7199 & 0.7116 & 0.6963 & 0.6984 & 0.6966 & 0.7254 \\
MageNet Baseline & 0.6908 & 0.6766 & 0.6984 & 0.6773 & 0.6738 & 0.6953 & 0.6895 & 0.7235 \\
MageNet Baseline ResGCN & 0.7034 & 0.7042 & 0.6980 & 0.6749 & 0.6784 & 0.6874 & 0.7076 & 0.6830 \\
GraphMAGE-SL ResGCN & 0.6974 & 0.7198 & 0.6999 & 0.7157 & 0.7000 & 0.7088 & 0.6960 & 0.7334 \\
Random Forest & 0.8279 & 0.8276 & 0.8257 & 0.8247 & 0.8275 & 0.8344 & 0.8376 & 0.8451 \\
Logistic Regression & 0.4485 & 0.4769 & 0.4659 & 0.4854 & 0.4707 & 0.5077 & 0.512 & 0.5199 \\
\midrule
\multicolumn{9}{c}{\textbf{Brier Score ($\downarrow$)}} \\
\midrule
Model & Week 0 & Week 1 & Week 2 & Week 3 & Week 4 & Week 5 & Week 6 & Week 7 \\
\midrule
GraphMAGE-SSL & 0.1753 & 0.1776 & 0.1882 & 0.189 & 0.1858 & 0.1828 & 0.183 & 0.1687 \\
GraphMAGE-SL & 0.1752 & 0.1843 & 0.1826 & 0.1842 & 0.1894 & 0.1863 & 0.1808 & 0.1658 \\
MageNet Baseline & 0.2019 & 0.2108 & 0.1996 & 0.2066 & 0.2063 & 0.1967 & 0.1983 & 0.1799 \\
MageNet Baseline ResGCN & 0.2265 & 0.2294 & 0.2260 & 0.2558 & 0.2520 & 0.2269 & 0.2178 & 0.2493 \\
GraphMAGE-SL ResGCN & 0.1920 & 0.1837 & 0.1902 & 0.1827 & 0.1893 & 0.1807 & 0.1793 & 0.1612 \\
Random Forest & 0.1204 & 0.1227 & 0.1245 & 0.1262 & 0.1272 & 0.1247 & 0.1215 & 0.1148 \\
Logistic Regression & 0.3349 & 0.326 & 0.3344 & 0.3204 & 0.331 & 0.2934 & 0.3053 & 0.2889 \\
\midrule
\multicolumn{9}{c}{\textbf{F1 Score ($\uparrow$)}} \\
\midrule
Model & Week 0 & Week 1 & Week 2 & Week 3 & Week 4 & Week 5 & Week 6 & Week 7 \\
\midrule
GraphMAGE-SSL & 0.561 & 0.5058 & 0.4864 & 0.4792 & 0.4717 & 0.4672 & 0.4659 & 0.4824 \\
GraphMAGE-SL & 0.5615 & 0.5033 & 0.4923 & 0.4833 & 0.4675 & 0.4622 & 0.4659 & 0.484 \\
MageNet Baseline & 0.433 & 0.3724 & 0.372 & 0.3599 & 0.3528 & 0.3534 & 0.3526 & 0.3468 \\
MageNet Baseline ResGCN & 0.4130 & 0.3251 & 0.3375 & 0.3239 & 0.3143 & 0.3384 & 0.3037 & 0.3103 \\
GraphMAGE-SL ResGCN & 0.4961 & 0.4452 & 0.4520 & 0.4516 & 0.4396 & 0.4382 & 0.4364 & 0.4358 \\
Random Forest & 0.3505 & 0.1971 & 0.1077 & 0.0397 & 0.0098 & 0.0002 & 0.0007 & 0 \\
Logistic Regression & 0.2986 & 0.2677 & 0.2544 & 0.2592 & 0.244 & 0.2607 & 0.2473 & 0.2354 \\
\midrule
\multicolumn{9}{c}{\textbf{Specificity ($\uparrow$)}} \\
\midrule
Model & Week 0 & Week 1 & Week 2 & Week 3 & Week 4 & Week 5 & Week 6 & Week 7 \\
\midrule
GraphMAGE-SSL & 0.72 & 0.724 & 0.6937 & 0.6851 & 0.6968 & 0.6959 & 0.673 & 0.6993 \\
GraphMAGE-SL& 0.7182 & 0.7054 & 0.7115 & 0.6991 & 0.6806 & 0.6819 & 0.6742 & 0.7066 \\
MageNet Baseline & 0.6907 & 0.6847 & 0.717 & 0.685 & 0.6817 & 0.7076 & 0.6992 & 0.7417 \\
MageNet Baseline ResGCN & 0.7200 & 0.7421 & 0.7270 & 0.6971 & 0.7062 & 0.7053 & 0.7387 & 0.7014 \\
GraphMAGE-SL ResGCN & 0.6762 & 0.7258 & 0.6914 & 0.7163 & 0.6942 & 0.7043 & 0.6812 & 0.7287 \\
Random Forest & 0.9822 & 0.9886 & 0.9922 & 0.9949 & 0.9985 & 0.9997 & 0.9999 & 1 \\
Logistic Regression & 0.4166 & 0.4679 & 0.4567 & 0.4799 & 0.4661 & 0.5045 & 0.5159 & 0.528 \\
\midrule
\multicolumn{9}{c}{\textbf{Sensitivity ($\uparrow$)}} \\
\midrule
Model & Week 0 & Week 1 & Week 2 & Week 3 & Week 4 & Week 5 & Week 6 & Week 7 \\
\midrule
GraphMAGE-SSL & 0.8135 & 0.7506 & 0.7719 & 0.7828 & 0.7564 & 0.7711 & 0.8142 & 0.8372 \\
GraphMAGE-SL & 0.8165 & 0.772 & 0.7587 & 0.7704 & 0.7715 & 0.7813 & 0.8122 & 0.8283 \\
MageNet Baseline & 0.6909 & 0.6407 & 0.6126 & 0.6408 & 0.6361 & 0.6332 & 0.6394 & 0.6245 \\
MageNet Baseline ResGCN & 0.6390 & 0.5367 & 0.5647 & 0.5700 & 0.5446 & 0.5970 & 0.5470 & 0.5824 \\
GraphMAGE-SL ResGCN & 0.7798 & 0.6934 & 0.7388 & 0.7129 & 0.7278 & 0.7315 & 0.7722 & 0.7591 \\
Random Forest & 0.2273 & 0.1149 & 0.059 & 0.0208 & 0.005 & 0.0001 & 0.0004 & 0 \\
Logistic Regression & 0.5728 & 0.5168 & 0.5081 & 0.5113 & 0.4925 & 0.5239 & 0.492 & 0.4757 \\
\hline
\end{tabular}
}
\caption{Overall - Trained On Upper 80 Model Evaluation Metrics by Week}
\label{tab:metrics_trainedOnUpper80_weeks_appendix}
\end{table}

%Upper 20

\begin{table}[ht!]
\scriptsize
\centering
\resizebox{\columnwidth}{!}{
\begin{tabular}{l|cccccccc}
\toprule
\multicolumn{9}{c}{\textbf{AUC Score ($\uparrow$)}} \\
\midrule
Model & Week 0 & Week 1 & Week 2 & Week 3 & Week 4 & Week 5 & Week 6 & Week 7 \\
\midrule
GraphMAGE-SSL & 0.8716 & 0.8392 & 0.8444 & 0.8383 & 0.8284 & 0.8342 & 0.8504 & 0.8725 \\
GraphMAGE-SL & 0.8729 & 0.8385 & 0.843 & 0.8358 & 0.8236 & 0.831 & 0.8474 & 0.8686 \\
MageNet Baseline & 0.8152 & 0.7921 & 0.7957 & 0.7884 & 0.7768 & 0.7925 & 0.7857 & 0.7981 \\
MageNet Baseline ResGCN & 0.7558 & 0.7190 & 0.7280 & 0.6891 & 0.6859 & 0.7298 & 0.7224 & 0.7372 \\
GraphMAGE-SL ResGCN & 0.8170 & 0.7828 & 0.7963 & 0.7903 & 0.7846 & 0.7913 & 0.8023 & 0.8348 \\
Random Forest & 0.8488 & 0.8081 & 0.7973 & 0.7788 & 0.7682 & 0.7744 & 0.7879 & 0.8093 \\
Logistic Regression & 0.483 & 0.4858 & 0.4641 & 0.489 & 0.4625 & 0.5284 & 0.504 & 0.5032 \\
\midrule
\multicolumn{9}{c}{\textbf{Accuracy ($\uparrow$)}} \\
\midrule
GraphMAGE-SSL & 0.7774 & 0.7562 & 0.7346 & 0.7197 & 0.7165 & 0.7217 & 0.7141 & 0.7456 \\
GraphMAGE-SL & 0.7754 & 0.7421 & 0.7414 & 0.7231 & 0.7038 & 0.7089 & 0.7114 & 0.7422 \\
MageNet Baseline & 0.7107 & 0.6948 & 0.7194 & 0.6923 & 0.6816 & 0.7077 & 0.7033 & 0.7311 \\
MageNet Baseline ResGCN & 0.7169 & 0.7061 & 0.7056 & 0.6709 & 0.6729 & 0.6778 & 0.7002 & 0.6838 \\
GraphMAGE-SL ResGCN & 0.7097 & 0.7264 & 0.7076 & 0.7094 & 0.6896 & 0.6919 & 0.6914 & 0.7292 \\
Random Forest & 0.8019 & 0.7879 & 0.7923 & 0.7845 & 0.7924 & 0.7992 & 0.8036 & 0.813 \\
Logistic Regression & 0.4559 & 0.4784 & 0.4659 & 0.4864 & 0.4698 & 0.511 & 0.5099 & 0.5171 \\
\midrule
\multicolumn{9}{c}{\textbf{Brier Score ($\downarrow$)}} \\
\midrule
GraphMAGE-SSL & 0.1569 & 0.167 & 0.1774 & 0.181 & 0.1816 & 0.1774 & 0.1747 & 0.1562 \\
GraphMAGE-SL & 0.1581 & 0.1757 & 0.1752 & 0.1786 & 0.188 & 0.184 & 0.1744 & 0.1584 \\
MageNet Baseline & 0.1861 & 0.1987 & 0.1887 & 0.1984 & 0.201 & 0.1909 & 0.1915 & 0.1752 \\
MageNet Baseline ResGCN & 0.2180 & 0.2329 & 0.2250 & 0.2596 & 0.2552 & 0.2324 & 0.2246 & 0.2509 \\
GraphMAGE-SL ResGCN & 0.1911 & 0.1909 & 0.1918 & 0.1912 & 0.1990 & 0.1911 & 0.1864 & 0.1704 \\
Random Forest & 0.1347 & 0.144 & 0.1429 & 0.149 & 0.1477 & 0.1467 & 0.1411 & 0.1331 \\
Logistic Regression & 0.3316 & 0.3253 & 0.3348 & 0.3212 & 0.333 & 0.2903 & 0.3074 & 0.2892 \\
\midrule
\multicolumn{9}{c}{\textbf{F1 Score ($\uparrow$)}} \\
\midrule
GraphMAGE-SSL & 0.6503 & 0.5941 & 0.5675 & 0.5566 & 0.5359 & 0.5375 & 0.5428 & 0.565 \\
GraphMAGE-SL & 0.6495 & 0.5882 & 0.5716 & 0.5549 & 0.5292 & 0.5322 & 0.5394 & 0.5612 \\
MageNet Baseline & 0.4941 & 0.4395 & 0.4359 & 0.4172 & 0.4064 & 0.4097 & 0.4121 & 0.3996 \\
MageNet Baseline ResGCN & 0.4722 & 0.3796 & 0.3907 & 0.3684 & 0.3548 & 0.3841 & 0.3466 & 0.3540 \\
GraphMAGE-SL ResGCN & 0.5545 & 0.5092 & 0.5077 & 0.4988 & 0.4779 & 0.4752 & 0.4876 & 0.4912 \\
Random Forest & 0.4174 & 0.2428 & 0.1337 & 0.0406 & 0.0042 & 0 & 0 & 0 \\
Logistic Regression & 0.3403 & 0.3101 & 0.2848 & 0.2986 & 0.2722 & 0.3044 & 0.2818 & 0.2697 \\
\midrule
\multicolumn{9}{c}{\textbf{Specificity ($\uparrow$)}} \\
\midrule
GraphMAGE-SSL & 0.7588 & 0.7523 & 0.7142 & 0.6922 & 0.697 & 0.7012 & 0.6779 & 0.7147 \\
GraphMAGE-SL & 0.7551 & 0.7271 & 0.7245 & 0.7008 & 0.6774 & 0.6807 & 0.676 & 0.7111 \\
MageNet Baseline & 0.7169 & 0.7106 & 0.7457 & 0.7064 & 0.6868 & 0.7229 & 0.7129 & 0.7511 \\
MageNet Baseline ResGCN  & 0.7391 & 0.7534 & 0.7383 & 0.6945 & 0.7006 & 0.6900 & 0.7301 & 0.7025 \\
GraphMAGE-SL ResGCN & 0.6849 & 0.7332 & 0.6953 & 0.7062 & 0.6769 & 0.6790 & 0.6642 & 0.7110 \\
Random Forest & 0.9712 & 0.9811 & 0.9878 & 0.9925 & 0.9973 & 1 & 0.9998 & 1 \\
Logistic Regression & 0.4192 & 0.47 & 0.4584 & 0.4812 & 0.4676 & 0.5057 & 0.5154 & 0.5266 \\
\midrule
\multicolumn{9}{c}{\textbf{Sensitivity ($\uparrow$)}} \\
\midrule
GraphMAGE-SSL & 0.8341 & 0.7692 & 0.8096 & 0.8208 & 0.7919 & 0.8032 & 0.8624 & 0.8799 \\
GraphMAGE-SL & 0.8371 & 0.7919 & 0.8035 & 0.8048 & 0.8053 & 0.821 & 0.8568 & 0.8778 \\
MageNet Baseline & 0.692 & 0.6425 & 0.6232 & 0.6407 & 0.6616 & 0.6468 & 0.6641 & 0.644 \\
MageNet Baseline ResGCN & 0.6493 & 0.5490 & 0.5855 & 0.5842 & 0.5662 & 0.6291 & 0.5775 & 0.6026 \\
GraphMAGE-SL ResGCN & 0.7849 & 0.7038 & 0.7529 & 0.7213 & 0.7386 & 0.7434 & 0.8027 & 0.8085 \\
Random Forest & 0.2869 & 0.1469 & 0.0749 & 0.0213 & 0.0021 & 0 & 0 & 0 \\
Logistic Regression & 0.5674 & 0.5062 & 0.4935 & 0.5059 & 0.4782 & 0.5319 & 0.4872 & 0.4759 \\
\bottomrule
\end{tabular}
}
\caption{Upper 20 percent - Trained On Upper 80 Model Evaluation Metrics by Week}
\label{tab:Upper20_metrics_trainedOnUpper80_weeks_appendix}
\end{table}

%Lower 20

\begin{table}[ht!]
\scriptsize
\centering
\resizebox{\columnwidth}{!}{
\begin{tabular}{l|cccccccc}
\toprule
\multicolumn{9}{c}{\textbf{AUC Score ($\uparrow$)}} \\
\midrule
Model & Week 0 & Week 1 & Week 2 & Week 3 & Week 4 & Week 5 & Week 6 & Week 7 \\
\midrule
GraphMAGE-SSL & 0.7621 & 0.7232 & 0.7008 & 0.6754 & 0.6879 & 0.694 & 0.716 & 0.7484 \\
GraphMAGE-SL & 0.7653 & 0.725 & 0.7079 & 0.6784 & 0.6924 & 0.6949 & 0.72 & 0.7497 \\
MageNet Baseline & 0.7148 & 0.6809 & 0.6683 & 0.6424 & 0.6406 & 0.6712 & 0.6755 & 0.7089 \\
MageNet Baseline ResGCN & 0.6584 & 0.6244 & 0.6188 & 0.5948 & 0.6052 & 0.6240 & 0.6426 & 0.6572 \\
GraphMAGE-SL ResGCN & 0.7219 & 0.6879 & 0.6757 & 0.6558 & 0.6681 & 0.6854 & 0.6994 & 0.7252 \\
Random Forest & 0.7801 & 0.7374 & 0.7214 & 0.6933 & 0.7026 & 0.6945 & 0.7137 & 0.7422 \\
Logistic Regression & 0.4953 & 0.4996 & 0.4815 & 0.5005 & 0.4959 & 0.509 & 0.5127 & 0.5012 \\
\midrule
\multicolumn{9}{c}{\textbf{Accuracy ($\uparrow$)}} \\
\midrule
GraphMAGE-SSL & 0.6613 & 0.6955 & 0.6733 & 0.6705 & 0.6854 & 0.6805 & 0.6697 & 0.6828 \\
GraphMAGE-SL & 0.6631 & 0.6865 & 0.6966 & 0.6837 & 0.6718 & 0.6731 & 0.6661 & 0.6975 \\
MageNet Baseline & 0.6479 & 0.6416 & 0.6629 & 0.6438 & 0.6516 & 0.6652 & 0.6602 & 0.7178 \\
MageNet Baseline ResGCN & 0.6702 & 0.7022 & 0.6933 & 0.6799 & 0.6875 & 0.7035 & 0.7282 & 0.6802 \\
GraphMAGE-SL ResGCN & 0.6490 & 0.7069 & 0.6873 & 0.7134 & 0.6979 & 0.7144 & 0.6927 & 0.7469 \\
Random Forest & 0.8899 & 0.9004 & 0.899 & 0.9043 & 0.9055 & 0.9094 & 0.9132 & 0.9219 \\
Logistic Regression & 0.4321 & 0.4791 & 0.4643 & 0.4823 & 0.4766 & 0.5006 & 0.5137 & 0.5235 \\
\midrule
\multicolumn{9}{c}{\textbf{Brier Score ($\downarrow$)}} \\
\midrule
GraphMAGE-SSL & 0.2198 & 0.1927 & 0.203 & 0.2035 & 0.1953 & 0.1941 & 0.1947 & 0.1861 \\
GraphMAGE-SL & 0.2169 & 0.1981 & 0.1921 & 0.1977 & 0.1986 & 0.1962 & 0.1929 & 0.1786 \\
MageNet Baseline & 0.2385 & 0.236 & 0.2222 & 0.2298 & 0.224 & 0.2144 & 0.2153 & 0.1881 \\
MageNet Baseline ResGCN & 0.2494 & 0.2268 & 0.2263 & 0.2535 & 0.2485 & 0.2197 & 0.2028 & 0.2510 \\
GraphMAGE-SL ResGCN & 0.2046 & 0.1745 & 0.1863 & 0.1738 & 0.1786 & 0.1692 & 0.1704 & 0.1443 \\
Random Forest & 0.0907 & 0.0848 & 0.0868 & 0.086 & 0.0856 & 0.0829 & 0.08 & 0.073 \\
Logistic Regression & 0.3408 & 0.3242 & 0.3329 & 0.3182 & 0.3248 & 0.2971 & 0.3035 & 0.2862 \\
\midrule
\multicolumn{9}{c}{\textbf{F1 Score ($\uparrow$)}} \\
\midrule
GraphMAGE-SSL & 0.3438 & 0.287 & 0.2688 & 0.2468 & 0.2496 & 0.2421 & 0.2477 & 0.2511 \\
GraphMAGE-SL & 0.3451 & 0.2891 & 0.2781 & 0.2538 & 0.2499 & 0.2429 & 0.2475 & 0.2522 \\
MageNet Baseline & 0.266 & 0.2078 & 0.209 & 0.1874 & 0.1796 & 0.1871 & 0.1854 & 0.1793 \\
MageNet Baseline ResGCN & 0.2466 & 0.1834 & 0.1819 & 0.1628 & 0.1567 & 0.1638 & 0.1545 & 0.1549 \\
GraphMAGE-SL ResGCN & 0.3030 & 0.2523 & 0.2564 & 0.2307 & 0.2306 & 0.2298 & 0.2353 & 0.2166 \\
Random Forest & 0.1818 & 0.0454 & 0.0014 & 0 & 0 & 0 & 0 & 0 \\
Logistic Regression & 0.1961 & 0.1707 & 0.1621 & 0.1595 & 0.1586 & 0.1564 & 0.1531 & 0.1327 \\
\midrule
\multicolumn{9}{c}{\textbf{Specificity ($\uparrow$)}} \\
\midrule
GraphMAGE-SSL & 0.6483 & 0.7048 & 0.6817 & 0.6814 & 0.699 & 0.6917 & 0.6739 & 0.683 \\
GraphMAGE-SL & 0.6503 & 0.692 & 0.7095 & 0.6962 & 0.6813 & 0.6822 & 0.6694 & 0.7012 \\
MageNet Baseline & 0.6494 & 0.6509 & 0.6778 & 0.6577 & 0.6692 & 0.6803 & 0.6724 & 0.7362 \\
MageNet Baseline ResGCN & 0.6843 & 0.7311 & 0.7221 & 0.7081 & 0.7183 & 0.7347 & 0.7610 & 0.6999 \\
GraphMAGE-SL ResGCN & 0.6432 & 0.7241 & 0.7007 & 0.7383 & 0.7180 & 0.7357 & 0.7038 & 0.7679 \\
Random Forest & 0.9933 & 0.9975 & 0.999 & 0.9994 & 0.9997 & 0.9995 & 1 & 1 \\
Logistic Regression & 0.411 & 0.473 & 0.4584 & 0.4788 & 0.4715 & 0.4992 & 0.514 & 0.5282 \\
\midrule
\multicolumn{9}{c}{\textbf{Sensitivity ($\uparrow$)}} \\
\midrule
GraphMAGE-SSL & 0.7603 & 0.612 & 0.5982 & 0.5667 & 0.5542 & 0.5669 & 0.6258 & 0.6808 \\
GraphMAGE-SL & 0.7599 & 0.6368 & 0.5804 & 0.5646 & 0.5801 & 0.5818 & 0.6319 & 0.6538 \\
MageNet Baseline & 0.6365 & 0.5579 & 0.5293 & 0.5117 & 0.4821 & 0.5128 & 0.5314 & 0.5009 \\
MageNet Baseline ResGCN & 0.5631 & 0.4407 & 0.4346 & 0.4117 & 0.3912 & 0.3885 & 0.3832 & 0.4476 \\
GraphMAGE-SL ResGCN & 0.6927 & 0.5512 & 0.5666 & 0.4769 & 0.5054 & 0.4995 & 0.5758 & 0.4988 \\
Random Forest & 0.1052 & 0.0238 & 0.0007 & 0 & 0 & 0 & 0 & 0 \\
Logistic Regression & 0.5917 & 0.5347 & 0.5171 & 0.516 & 0.5248 & 0.5149 & 0.5101 & 0.468 \\
\bottomrule
\end{tabular}
}
\caption{Lower 20 percent - Trained On Upper 80 Model Evaluation Metrics by Week}
\label{tab:Upper 20 metrics_trainedOnUpper80_weeks_appendix}
\end{table}

%Upper 20 percent held out 

%Overall

\begin{table}[ht!]
\scriptsize
\centering
\resizebox{\columnwidth}{!}{
\begin{tabular}{l|cccccccc}
\toprule

\multicolumn{9}{c}{\textbf{AUC Score ($\uparrow$)}} \\
\midrule
Model & Week 0 & Week 1 & Week 2 & Week 3 & Week 4 & Week 5 & Week 6 & Week 7 \\
\midrule

GraphMAGE-SSL & 0.8473 & 0.8165 & 0.8099 & 0.8069 & 0.8045 & 0.8141 & 0.8244 & 0.8447 \\
GraphMAGE-SL & 0.8467 & 0.8169 & 0.8091 & 0.808 & 0.8054 & 0.8113 & 0.8223 & 0.8461 \\
MageNet Baseline & 0.815 & 0.7852 & 0.7832 & 0.78 & 0.7815 & 0.7821 & 0.7986 & 0.819 \\
MageNet Baseline ResGCN & 0.7450 & 0.6920 & 0.6606 & 0.6594 & 0.6612 & 0.6494 & 0.6703 & 0.6908 \\
GraphMAGE-SL ResGCN & 0.7965 & 0.7998 & 0.7775 & 0.7768 & 0.7752 & 0.7911 & 0.8144 & 0.8448 \\
Random Forest & 0.8359 & 0.7972 & 0.78 & 0.7705 & 0.759 & 0.7616 & 0.7737 & 0.7909 \\
Logistic Regression & 0.4819 & 0.4861 & 0.4719 & 0.4936 & 0.473 & 0.5183 & 0.5045 & 0.5013 \\
\midrule
\multicolumn{9}{c}{\textbf{Accuracy ($\uparrow$)}} \\
\midrule
Model & Week 0 & Week 1 & Week 2 & Week 3 & Week 4 & Week 5 & Week 6 & Week 7 \\
\midrule
GraphMAGE-SSL & 0.7418 & 0.7556 & 0.7454 & 0.7439 & 0.7393 & 0.7353 & 0.7312 & 0.7324 \\
GraphMAGE-SL & 0.7448 & 0.7444 & 0.7323 & 0.7402 & 0.7316 & 0.739 & 0.7181 & 0.7334 \\
MageNet Baseline & 0.7442 & 0.7578 & 0.7615 & 0.7573 & 0.753 & 0.7497 & 0.7374 & 0.7539 \\
MageNet Baseline ResGCN & 0.7413 & 0.7560 & 0.7627 & 0.7616 & 0.7556 & 0.7650 & 0.7469 & 0.7706 \\
GraphMAGE-SL ResGCN & 0.7084 & 0.7138 & 0.7201 & 0.7307 & 0.7272 & 0.7260 & 0.7071 & 0.7202 \\
Random Forest & 0.8249 & 0.8249 & 0.8239 & 0.8251 & 0.8279 & 0.8346 & 0.8376 & 0.8451 \\
Logistic Regression & 0.5165 & 0.5515 & 0.541 & 0.5648 & 0.5497 & 0.6003 & 0.5875 & 0.6103 \\
\midrule
\multicolumn{9}{c}{\textbf{Brier Score ($\downarrow$)}} \\
\midrule
Model & Week 0 & Week 1 & Week 2 & Week 3 & Week 4 & Week 5 & Week 6 & Week 7 \\
\midrule
GraphMAGE-SSL & 0.1689 & 0.1633 & 0.1667 & 0.1692 & 0.1701 & 0.167 & 0.1653 & 0.1597 \\
GraphMAGE-SL & 0.1677 & 0.1697 & 0.1758 & 0.17 & 0.1725 & 0.1666 & 0.1715 & 0.158 \\
MageNet Baseline & 0.167 & 0.1638 & 0.162 & 0.1627 & 0.1652 & 0.1637 & 0.161 & 0.1539 \\
MageNet Baseline ResGCN & 0.1855 & 0.1820 & 0.1800 & 0.1801 & 0.1829 & 0.1737 & 0.1838 & 0.1626 \\
GraphMAGE-SL ResGCN & 0.1882 & 0.1819 & 0.1797 & 0.1754 & 0.1752 & 0.1723 & 0.1738 & 0.1611 \\
Random Forest & 0.1235 & 0.125 & 0.1273 & 0.1297 & 0.131 & 0.1281 & 0.1252 & 0.119 \\
Logistic Regression & 0.295 & 0.2841 & 0.2902 & 0.2767 & 0.2843 & 0.2512 & 0.2626 & 0.245 \\
\midrule
\multicolumn{9}{c}{\textbf{F1 Score ($\uparrow$)}} \\
\midrule
Model & Week 0 & Week 1 & Week 2 & Week 3 & Week 4 & Week 5 & Week 6 & Week 7 \\
\midrule
GraphMAGE-SSL & 0.5605 & 0.5187 & 0.4975 & 0.489 & 0.4776 & 0.474 & 0.4789 & 0.4841 \\
GraphMAGE-SL & 0.5618 & 0.5145 & 0.4942 & 0.4904 & 0.4776 & 0.4724 & 0.4689 & 0.4856 \\
MageNet Baseline & 0.5116 & 0.4098 & 0.3869 & 0.3831 & 0.3637 & 0.3715 & 0.3894 & 0.3762 \\
MageNet Baseline ResGCN & 0.4229 & 0.3184 & 0.2642 & 0.2754 & 0.2628 & 0.2614 & 0.2492 & 0.2747 \\
GraphMAGE-SL ResGCN & 0.4975 & 0.4917 & 0.4543 & 0.4501 & 0.4395 & 0.4408 & 0.4629 & 0.4804 \\
Random Forest & 0.3079 & 0.1611 & 0.0727 & 0.0121 & 0.0011 & 0 & 0 & 0 \\
Logistic Regression & 0.2685 & 0.2444 & 0.2311 & 0.2418 & 0.2288 & 0.2434 & 0.2361 & 0.2196 \\
\midrule
\multicolumn{9}{c}{\textbf{Specificity ($\uparrow$)}} \\
\midrule
Model & Week 0 & Week 1 & Week 2 & Week 3 & Week 4 & Week 5 & Week 6 & Week 7 \\
\midrule
GraphMAGE-SSL & 0.7264 & 0.7653 & 0.7543 & 0.753 & 0.7494 & 0.7383 & 0.7262 & 0.7186 \\
GraphMAGE-SL & 0.7313 & 0.7473 & 0.7338 & 0.7458 & 0.7359 & 0.7457 & 0.7096 & 0.7195 \\
MageNet Baseline & 0.7558 & 0.8034 & 0.813 & 0.8058 & 0.8038 & 0.7923 & 0.7607 & 0.7804 \\
MageNet Baseline ResGCN & 0.7784 & 0.8232 & 0.8462 & 0.8367 & 0.8302 & 0.8389 & 0.8119 & 0.8288 \\
GraphMAGE-SL ResGCN & 0.6966 & 0.7059 & 0.7260 & 0.7434 & 0.7413 & 0.7324 & 0.6939 & 0.6996 \\
Random Forest & 0.9878 & 0.9907 & 0.9944 & 0.9984 & 0.9999 & 1 & 1 & 1 \\
Logistic Regression & 0.5375 & 0.5878 & 0.5748 & 0.6009 & 0.5838 & 0.642 & 0.6255 & 0.6573 \\
\midrule
\multicolumn{9}{c}{\textbf{Sensitivity ($\uparrow$)}} \\
\midrule
Model & Week 0 & Week 1 & Week 2 & Week 3 & Week 4 & Week 5 & Week 6 & Week 7 \\
\midrule
GraphMAGE-SSL & 0.8021 & 0.7124 & 0.7045 & 0.7005 & 0.6902 & 0.72 & 0.757 & 0.8075 \\
GraphMAGE-SL & 0.7975 & 0.7317 & 0.7253 & 0.7136 & 0.7109 & 0.7055 & 0.762 & 0.809 \\
MageNet Baseline & 0.6989 & 0.556 & 0.5245 & 0.5279 & 0.5092 & 0.5348 & 0.6174 & 0.6096 \\
MageNet Baseline ResGCN & 0.5970 & 0.4586 & 0.3783 & 0.4072 & 0.3969 & 0.3921 & 0.4113 & 0.4532 \\
GraphMAGE-SL ResGCN & 0.7543 & 0.7485 & 0.6925 & 0.6703 & 0.6595 & 0.6938 & 0.7751 & 0.8323 \\
Random Forest & 0.1906 & 0.0912 & 0.0387 & 0.0061 & 0.0005 & 0 & 0 & 0 \\
Logistic Regression & 0.4346 & 0.3912 & 0.3853 & 0.394 & 0.386 & 0.3902 & 0.3914 & 0.354 \\
\hline
\end{tabular}
}
\caption{Overall - Trained On Lower 80 Model Evaluation Metrics by Week}
\label{tab:metrics_trainedOnLower80_weeks_appendix}
\end{table}

%Upper 20

\begin{table}[ht!]
\scriptsize
\centering
\resizebox{\columnwidth}{!}{
\begin{tabular}{l|cccccccc}
\toprule

\multicolumn{9}{c}{\textbf{AUC Score ($\uparrow$)}} \\
\midrule
Model & Week 0 & Week 1 & Week 2 & Week 3 & Week 4 & Week 5 & Week 6 & Week 7 \\
\midrule

GraphMAGE-SSL & 0.8657 & 0.8341 & 0.8384 & 0.82 & 0.8147 & 0.8263 & 0.8437 & 0.8598 \\
GraphMAGE-SL & 0.8654 & 0.8338 & 0.8346 & 0.8193 & 0.8156 & 0.8195 & 0.8379 & 0.8592 \\
MageNet Baseline & 0.8237 & 0.789 & 0.7956 & 0.7807 & 0.782 & 0.7809 & 0.8054 & 0.8279 \\
MageNet Baseline ResGCN & 0.7581 & 0.7018 & 0.6683 & 0.6611 & 0.6595 & 0.6516 & 0.6738 & 0.6952 \\
GraphMAGE-SL ResGCN & 0.8019 & 0.7992 & 0.7874 & 0.7757 & 0.7697 & 0.7890 & 0.8066 & 0.8376 \\
Random Forest & 0.8406 & 0.8004 & 0.7837 & 0.767 & 0.756 & 0.7606 & 0.7843 & 0.8009 \\
Logistic Regression & 0.4775 & 0.4822 & 0.4633 & 0.4907 & 0.465 & 0.5238 & 0.5015 & 0.5002 \\
\midrule
\multicolumn{9}{c}{\textbf{Accuracy ($\uparrow$)}} \\
\midrule
Model & Week 0 & Week 1 & Week 2 & Week 3 & Week 4 & Week 5 & Week 6 & Week 7 \\
\midrule
GraphMAGE-SSL & 0.7626 & 0.7642 & 0.7558 & 0.7408 & 0.7297 & 0.7363 & 0.7412 & 0.7425 \\
GraphMAGE-SL & 0.7705 & 0.7528 & 0.738 & 0.7348 & 0.7238 & 0.7309 & 0.7221 & 0.7379 \\
MageNet Baseline & 0.7379 & 0.746 & 0.7572 & 0.7449 & 0.7384 & 0.7351 & 0.7315 & 0.7533 \\
MageNet Baseline ResGCN & 0.7372 & 0.7402 & 0.7479 & 0.7416 & 0.7355 & 0.7464 & 0.7259 & 0.7568 \\
GraphMAGE-SL ResGCN & 0.6997 & 0.7093 & 0.7081 & 0.7105 & 0.6971 & 0.6977 & 0.6929 & 0.7186 \\
Random Forest & 0.7973 & 0.7829 & 0.7898 & 0.7854 & 0.7938 & 0.7992 & 0.8038 & 0.813 \\
Logistic Regression & 0.5087 & 0.5416 & 0.5304 & 0.5565 & 0.5408 & 0.5954 & 0.5793 & 0.5997 \\
\midrule
\multicolumn{9}{c}{\textbf{Brier Score ($\downarrow$)}} \\
\midrule
Model & Week 0 & Week 1 & Week 2 & Week 3 & Week 4 & Week 5 & Week 6 & Week 7 \\
\midrule
GraphMAGE-SSL & 0.1628 & 0.1621 & 0.1631 & 0.1709 & 0.1737 & 0.1683 & 0.1622 & 0.1558 \\
GraphMAGE-SL & 0.1596 & 0.1689 & 0.1748 & 0.1729 & 0.1762 & 0.1713 & 0.172 & 0.1568 \\
MageNet Baseline & 0.1674 & 0.169 & 0.1635 & 0.1685 & 0.17 & 0.1689 & 0.1621 & 0.1539 \\
MageNet Baseline ResGCN & 0.1904 & 0.1966 & 0.1946 & 0.1985 & 0.1994 & 0.1885 & 0.2000 & 0.1738 \\
GraphMAGE-SL ResGCN & 0.1965 & 0.1907 & 0.1893 & 0.1893 & 0.1893 & 0.1859 & 0.1823 & 0.1667 \\
Random Forest & 0.1405 & 0.1494 & 0.1487 & 0.1557 & 0.1545 & 0.153 & 0.1477 & 0.1408 \\
Logistic Regression & 0.2982 & 0.2902 & 0.2957 & 0.2824 & 0.2908 & 0.253 & 0.269 & 0.2493 \\
\midrule
\multicolumn{9}{c}{\textbf{F1 Score ($\uparrow$)}} \\
\midrule
Model & Week 0 & Week 1 & Week 2 & Week 3 & Week 4 & Week 5 & Week 6 & Week 7 \\
\midrule
GraphMAGE-SSL & 0.6365 & 0.5936 & 0.5692 & 0.5461 & 0.5225 & 0.5383 & 0.5507 & 0.5551 \\
GraphMAGE-SL & 0.6417 & 0.5882 & 0.5607 & 0.5457 & 0.525 & 0.5259 & 0.5353 & 0.5509 \\
MageNet Baseline & 0.5593 & 0.4549 & 0.4341 & 0.4251 & 0.3935 & 0.4091 & 0.4376 & 0.4265 \\
MageNet Baseline ResGCN & 0.4709 & 0.3588 & 0.2920 & 0.3039 & 0.2857 & 0.2885 & 0.2799 & 0.3057 \\
GraphMAGE-SL ResGCN & 0.5448 & 0.5524 & 0.5010 & 0.4886 & 0.4620 & 0.4732 & 0.5078 & 0.5406 \\
Random Forest & 0.3705 & 0.1976 & 0.0864 & 0.0087 & 0 & 0 & 0 & 0 \\
Logistic Regression & 0.2958 & 0.2808 & 0.253 & 0.2777 & 0.2521 & 0.2831 & 0.2679 & 0.2445 \\

\midrule
\multicolumn{9}{c}{\textbf{Specificity ($\uparrow$)}} \\
\midrule
Model & Week 0 & Week 1 & Week 2 & Week 3 & Week 4 & Week 5 & Week 6 & Week 7 \\
\midrule
GraphMAGE-SL & 0.7382 & 0.7706 & 0.7568 & 0.7443 & 0.7328 & 0.7293 & 0.7258 & 0.7165 \\
GraphMAGE-SL & 0.7518 & 0.7505 & 0.728 & 0.7329 & 0.7192 & 0.7278 & 0.6997 & 0.7108 \\
MageNet Baseline & 0.7445 & 0.7989 & 0.8157 & 0.8011 & 0.7962 & 0.7818 & 0.7529 & 0.7801 \\
MageNet Baseline ResGCN & 0.7772 & 0.8215 & 0.8460 & 0.8314 & 0.8216 & 0.8336 & 0.7974 & 0.8233 \\
GraphMAGE-SL ResGCN & 0.6734 & 0.6901 & 0.7026 & 0.7164 & 0.7034 & 0.6914 & 0.6643 & 0.6807 \\
Random Forest & 0.9801 & 0.9841 & 0.9924 & 0.9982 & 0.9997 & 1 & 1 & 1 \\
Logistic Regression & 0.5378 & 0.5886 & 0.5741 & 0.6007 & 0.5838 & 0.6445 & 0.6255 & 0.6579 \\
\midrule
\multicolumn{9}{c}{\textbf{Sensitivity ($\uparrow$)}} \\
\midrule
Model & Week 0 & Week 1 & Week 2 & Week 3 & Week 4 & Week 5 & Week 6 & Week 7 \\
\midrule
GraphMAGE-SSL & 0.837 & 0.743 & 0.7518 & 0.7281 & 0.7179 & 0.764 & 0.8043 & 0.8558 \\
GraphMAGE-SL & 0.8275 & 0.7602 & 0.7744 & 0.742 & 0.7415 & 0.7434 & 0.8138 & 0.8558 \\
MageNet Baseline & 0.7181 & 0.5701 & 0.5425 & 0.5388 & 0.5154 & 0.5495 & 0.6441 & 0.6364 \\
MageNet Baseline ResGCN & 0.6154 & 0.4704 & 0.3877 & 0.4120 & 0.4034 & 0.3997 & 0.4329 & 0.4677 \\
GraphMAGE-SL ResGCN & 0.7796 & 0.7729 & 0.7283 & 0.6889 & 0.6730 & 0.7226 & 0.8101 & 0.8836 \\
Random Forest & 0.2412 & 0.1155 & 0.0464 & 0.0044 & 0 & 0 & 0 & 0 \\
Logistic Regression & 0.4201 & 0.3857 & 0.37 & 0.3945 & 0.3746 & 0.4 & 0.3903 & 0.3467 \\
\hline
\end{tabular}
}
\caption{Upper 20 percent - Trained On Lower 80 Model Evaluation Metrics by Week}
\label{tab:upper20_metrics_trainedOnLower80_weeks_appendix}
\end{table}

%Lower 20

\begin{table}[ht!]
\scriptsize
\centering
\resizebox{\columnwidth}{!}{
\begin{tabular}{l|cccccccc}
\toprule

\multicolumn{9}{c}{\textbf{AUC Score ($\uparrow$)}} \\
\midrule
Model & Week 0 & Week 1 & Week 2 & Week 3 & Week 4 & Week 5 & Week 6 & Week 7 \\
\midrule

GraphMAGE-SSL & 0.7551 & 0.7237 & 0.6953 & 0.6779 & 0.6928 & 0.6984 & 0.7203 & 0.7511 \\
GraphMAGE-SL & 0.7539 & 0.7242 & 0.7011 & 0.6802 & 0.6919 & 0.6964 & 0.7196 & 0.755 \\
MageNet Baseline & 0.7274 & 0.7012 & 0.6916 & 0.664 & 0.6775 & 0.6858 & 0.7113 & 0.7403 \\
MageNet Baseline ResGCN & 0.6652 & 0.6280 & 0.6079 & 0.5941 & 0.6125 & 0.6026 & 0.6213 & 0.6507 \\
GraphMAGE-SL ResGCN & 0.6953 & 0.6832 & 0.6596 & 0.6517 & 0.6639 & 0.6778 & 0.7020 & 0.7352 \\
Random Forest & 0.7849 & 0.7395 & 0.7229 & 0.6981 & 0.7035 & 0.6918 & 0.7108 & 0.7389 \\
Logistic Regression & 0.4923 & 0.4966 & 0.48 & 0.5001 & 0.4965 & 0.5097 & 0.5127 & 0.5009 \\
\midrule
\multicolumn{9}{c}{\textbf{Accuracy ($\uparrow$)}} \\
\midrule
Model & Week 0 & Week 1 & Week 2 & Week 3 & Week 4 & Week 5 & Week 6 & Week 7 \\
\midrule
GraphMAGE-SSL & 0.7134 & 0.7693 & 0.7616 & 0.7685 & 0.7649 & 0.762 & 0.7505 & 0.7458 \\
GraphMAGE-SL & 0.7118 & 0.7585 & 0.756 & 0.7615 & 0.7478 & 0.7667 & 0.7309 & 0.744 \\
MageNet Baseline & 0.7702 & 0.8064 & 0.8113 & 0.7994 & 0.8035 & 0.7939 & 0.7732 & 0.7859 \\
MageNet Baseline ResGCN & 0.7574 & 0.8003 & 0.8119 & 0.8108 & 0.8090 & 0.8099 & 0.7948 & 0.8125 \\
GraphMAGE-SL ResGCN & 0.7197 & 0.7452 & 0.7674 & 0.7808 & 0.7793 & 0.7782 & 0.7419 & 0.7478 \\
Random Forest & 0.8884 & 0.9 & 0.8993 & 0.9048 & 0.9058 & 0.9099 & 0.9132 & 0.9219 \\
Logistic Regression & 0.526 & 0.5737 & 0.5607 & 0.5848 & 0.5727 & 0.6136 & 0.606 & 0.6329 \\
\midrule
\multicolumn{9}{c}{\textbf{Brier Score ($\downarrow$)}} \\
\midrule
Model & Week 0 & Week 1 & Week 2 & Week 3 & Week 4 & Week 5 & Week 6 & Week 7 \\
\midrule
GraphMAGE-SSL & 0.177 & 0.1518 & 0.1567 & 0.1559 & 0.1556 & 0.1521 & 0.153 & 0.149 \\
GraphMAGE-SL & 0.1796 & 0.1582 & 0.1613 & 0.158 & 0.1615 & 0.1508 & 0.1597 & 0.1478 \\
MageNet Baseline & 0.1558 & 0.1419 & 0.1399 & 0.1447 & 0.1461 & 0.1474 & 0.147 & 0.1437 \\
MageNet Baseline ResGCN & 0.1701 & 0.1478 & 0.1407 & 0.1390 & 0.1420 & 0.1379 & 0.1448 & 0.1300 \\
GraphMAGE-SL ResGCN & 0.1709 & 0.1577 & 0.1512 & 0.1455 & 0.1463 & 0.1432 & 0.1500 & 0.1408 \\
Random Forest & 0.0877 & 0.0823 & 0.0839 & 0.0826 & 0.0825 & 0.0803 & 0.0772 & 0.0695 \\
Logistic Regression & 0.2878 & 0.2699 & 0.277 & 0.2634 & 0.2674 & 0.2431 & 0.2501 & 0.232 \\
\midrule
\multicolumn{9}{c}{\textbf{F1 Score ($\uparrow$)}} \\
\midrule
Model & Week 0 & Week 1 & Week 2 & Week 3 & Week 4 & Week 5 & Week 6 & Week 7 \\
\midrule
GraphMAGE-SSL & 0.3436 & 0.2879 & 0.2547 & 0.2382 & 0.2418 & 0.2273 & 0.2445 & 0.2417 \\
GraphMAGE-SL & 0.3427 & 0.291 & 0.265 & 0.243 & 0.2479 & 0.2297 & 0.2376 & 0.2476 \\
MageNet Baseline & 0.3032 & 0.2317 & 0.2097 & 0.1817 & 0.1733 & 0.1781 & 0.2035 & 0.1937 \\
MageNet Baseline ResGCN & 0.2537 & 0.1743 & 0.1467 & 0.1362 & 0.1329 & 0.1236 & 0.1289 & 0.1379 \\
GraphMAGE-SL ResGCN & 0.2927 & 0.2611 & 0.2211 & 0.1957 & 0.1907 & 0.1914 & 0.2189 & 0.2236 \\
Random Forest & 0.112 & 0.0091 & 0 & 0 & 0 & 0 & 0 & 0 \\
Logistic Regression & 0.1833 & 0.1584 & 0.1513 & 0.1524 & 0.1496 & 0.1491 & 0.1484 & 0.1303 \\

\midrule
\multicolumn{9}{c}{\textbf{Specificity ($\uparrow$)}} \\
\midrule
Model & Week 0 & Week 1 & Week 2 & Week 3 & Week 4 & Week 5 & Week 6 & Week 7 \\
\midrule
GraphMAGE-SSL & 0.7226 & 0.8027 & 0.8005 & 0.809 & 0.8026 & 0.7985 & 0.7775 & 0.7646 \\
GraphMAGE-SL & 0.7206 & 0.7875 & 0.7907 & 0.7989 & 0.7794 & 0.804 & 0.7539 & 0.7609 \\
MageNet Baseline & 0.8123 & 0.8585 & 0.8684 & 0.8546 & 0.8599 & 0.8436 & 0.8102 & 0.8182 \\
MageNet Baseline ResGCN & 0.7998 & 0.8565 & 0.8759 & 0.8717 & 0.8694 & 0.8687 & 0.8449 & 0.8562 \\
GraphMAGE-SL ResGCN & 0.7451 & 0.7776 & 0.8138 & 0.8313 & 0.8295 & 0.8248 & 0.7720 & 0.7714 \\
Random Forest & 0.9974 & 0.9992 & 0.9995 & 1 & 1 & 1 & 1 & 1 \\
Logistic Regression & 0.5353 & 0.593 & 0.5794 & 0.6049 & 0.5905 & 0.6369 & 0.6256 & 0.6564 \\
\midrule
\multicolumn{9}{c}{\textbf{Sensitivity ($\uparrow$)}} \\
\midrule
Model & Week 0 & Week 1 & Week 2 & Week 3 & Week 4 & Week 5 & Week 6 & Week 7 \\
\midrule
GraphMAGE-SSL & 0.6437 & 0.4685 & 0.4131 & 0.3835 & 0.4029 & 0.3933 & 0.4662 & 0.5243 \\
GraphMAGE-SL & 0.6448 & 0.497 & 0.4438 & 0.4061 & 0.4444 & 0.39 & 0.4886 & 0.5444 \\
MageNet Baseline & 0.4508 & 0.3363 & 0.2986 & 0.2743 & 0.261 & 0.2923 & 0.3832 & 0.4044 \\
MageNet Baseline ResGCN & 0.4351 & 0.2935 & 0.2373 & 0.2308 & 0.2284 & 0.2164 & 0.2684 & 0.2970 \\
GraphMAGE-SL ResGCN & 0.5264 & 0.4530 & 0.3512 & 0.3010 & 0.2963 & 0.3082 & 0.4255 & 0.4680 \\
Random Forest & 0.0605 & 0.0046 & 0 & 0 & 0 & 0 & 0 & 0 \\
Logistic Regression & 0.456 & 0.4 & 0.3929 & 0.3937 & 0.4012 & 0.3779 & 0.3992 & 0.3553 \\
\hline
\end{tabular}
}
\caption{Lower 20 percent - Trained On Lower 80 Model Evaluation Metrics by Week}
\label{tab:lower20_metrics_trainedOnLower80_weeks_appendix}
\end{table}

\end{document}